\documentclass[a4paper,11pt]{article}

\usepackage[utf8]{inputenc}
\usepackage[english]{babel}

\usepackage{amsmath}
\usepackage{mathtools}
\usepackage[new]{old-arrows}

\usepackage{amsthm}
\usepackage{amsfonts}
\usepackage[section]{placeins}

\usepackage[font={small,it}]{caption}

\usepackage{float}

\usepackage{hyperref}

\usepackage{todonotes}

\usepackage{tikz}
\usetikzlibrary{matrix,arrows,decorations}

\newtheorem{theorem}{Theorem}[section]

\theoremstyle{remark}
\newtheorem{remark}[theorem]{Remark}

\theoremstyle{definition}
\newtheorem{definition}[theorem]{Definition}

\title{Denting the FRTB IMA computational challenge via Orthogonal Chebyshev Sliding Technique}

\author{Mariano Zeron-Medina Laris\footnote{m.zeron@iruiztechnologies.com}\\ Ignacio Ruiz\footnote{i.ruiz@iruiztechnologies.com}}

\date{\parbox{\linewidth}{\centering%
  \today\endgraf\bigskip\bigskip
  \large A version of this paper has been peer-reviewed and will be published in Wilmott Magazine in January $2021$.}}

\begin{document}
\maketitle

\begin{abstract}

In this paper we introduce a new technique based on high-dimensional Chebyshev Tensors called \emph{Orthogonal Chebyshev Sliding Technique}. We implemented this technique inside the systems of a tier-one bank to approximate Front Office pricing functions with the aim of reducing the substantial computational burden associated with the FRTB IMA capital calculation. In all cases, the computational burden reductions obtained were of more than $90\%$, while keeping high degrees of accuracy. The latter obtained as a result of the mathematical properties enjoyed by Chebyshev Tensors.

\end{abstract}

\section{Introduction}

The calculation of capital within the Fundamental Review of the Trading Book regulation (FRTB) requires , under IMA, the daily calculation of expected shortfalls (ES) with different liquidity horizons. The calculation of ES requires the pricing of portfolios on hundreds if not thousands of scenarios. This means each trade needs to be priced daily, depending on the different liquidity horizons, between 250 and a few thousand scenarios. On top of this daily calculation, there is a period of stress that must be estimated at least quarterly, which involves the valuation, of each trade, on around 3000 scenarios (around 10 years of historical data). Although the latter must be done quarterly, there is heavy pressure from regulators to do it monthly. The PLA test, which compares daily risk-theoretical P$\&$L with the daily Hypothetical P$\&$L for each trading desk, has forced banks to price their portfolios for the calculation of ES with Front Office systems. This considerably increases the operational and computational burden on banks forcing them to look for ways to reduce these costs.
\bigskip

In this paper we present a technique that reduces the computational burden that comes with pricing portfolios on hundreds or thousands of scenarios and keeps precision ranges well within what is required by regulators. The technique is called the \emph{Orthogonal Chebyshev sliding} technique. This technique is based on Chebyshev interpolants which have, amongst others, two unique mathematical properties. The first is that they converge exponentially to analytic functions. The second is the existence of algorithms that guarantee a fast and numerically stable evaluation of the Orthogonal Chebyshev Slider. These two properties, as will become clear in subsequent sections, make Chebyshev Sliders unique pricing function approximators as they massively reduce the computational burden associated with risk calculations while keeping high levels of accuracy.
\bigskip

In this paper we present the results obtained from a proof-of-concept (PoC) done within the systems of a tier-one bank in a period of 10 weeks. The aim was to compare the accuracy and time taken between the brute force approach  and the Orthogonal Chebyshev Sliding Technique, when computing the ES for different liquidity horizons, on a set of historical risk factors that span ten years and include a period of stress, within the FRTB IMA framework, for a portfolio of Interest Rate Swaps and a portfolio of Vanilla Swaptions. 
\bigskip

The agreed success criteria were that the computational burden should be reduced by at least $50\%$, while the accuracy at the level of ES should be of no more than $10\%$. All this while showing that the implementation of the Orthogonal Chebyshev Sliding technique within their systems could be achieved without much trouble. In addition to this, a range of different slider configurations were to be tested and their impact on the results assessed. A full description of the PoC and the way the technique was implemented is given in Section \ref{sec: PoC description}. The full set of results, where the reader can verify that the success criteria were met well beyond the expectations, are presented in Section \ref{sec: Results}.
\bigskip

The rest of the paper is organised as follows. In Section \ref{sec: cheb interpolants} we present the mathematical backbone of the Chebyshev Sliding Technique which is based on Chebyshev approximation theory. The main mathematical results on which the technique rests are presented both for the one-dimensional and the multi-dimensional case. In Section \ref{sec: subsection sliding technique} we see how the Chebyshev interpolants presented in Section \ref{sec: cheb interpolants} come together to form a Chebyshev Slider and how the latter is evaluated. In Section \ref{sec: subsection PCA} we see how the Chebyshev sliding technique is coupled with Principal Component Analysis (PCA) to produce Orthogonal Chebyshev Sliders and increase the power of the technique. In Section \ref{sec: PoC description} we describe the details of the case study carried out within the risk systems of the tier-one bank, where the Orthogonal Chebyshev Sliding Technique was implemented and applied to a portfolio of Swaps and a portfolio of Swaptions for the calculation of ES on a data set corresponding to ten years including a period of stress. Details of how the same slider is used for the calculation of ES on different liquidity horizons is also presented. In Section \ref{sec: Results} we present the results obtained. Finally, in Section \ref{sec: Key takeaways} we conclude with a brief summary of the results, their implications within capital calculations under the FRTB IMA framework, and future directions of research.

\section{Chebyshev interpolants}\label{sec: cheb interpolants}

In this Section we describe the theoretical background of the technique used to approximate the pricing functions used in the calculation of ES with different liquidity horizons under the FRTB IMA framework.
\bigskip

At the heart of the Chebyshev sliding technique we have Chebyshev interpolants and their mathematical properties. Next, we briefly cover the main results underpinning the power of Chebyshev interpolants as pricing function approximators. For more details we refer the reader to \cite{TrefethenTextbook},  and \cite{MoCaXChebUltra}.

\subsection{One-dimensional case}\label{sec: cheb interpolants dim one}

Polynomial interpolants are often thought as poor approximators. The bad reputation is owed in part to results that have been around for many decades. The first one is due to Runge who gave an example of an analytic function for which equidistant interpolation diverges exponentially \cite{Runge}. Analytic functions, by definition, enjoy a high degree of smoothness. Equidistant points are a natural choice for interpolation if there is no a-priori information to say otherwise. This example shows how polynomial interpolation, if not done properly, can have terrible consequences even on well behaved functions. 
\bigskip

The second result, which also goes back a long way, says that there is no interpolation scheme that guarantees convergence for the set of continuous functions \cite{Faber}. 
\bigskip

Results such as the ones mentioned above cemented a belief that using polynomial interpolants as approximators of functions (even analytic ones) is not appropriate. Even textbooks in the subject of function approximation warn against the use of polynomial interpolants (Appendix in \cite{TrefethenTextbook}). What is often missed, is that interpolation on carefully chosen distribution of points can yield optimal approximation properties if applied to the correct class of functions.
\bigskip

\begin{definition}
The Chebyshev points associated with the natural number $n$ are the real part of the points
\end{definition}

\begin{equation}
x_j = \mathrm{Re}(z_j) = \frac{1}{2}(z_j+z_j^{-1} ),\ \ \ \ \ 0\leq j \leq n.
\end{equation}
\bigskip

Equivalently, Chebyshev points can be defined as

\begin{equation}
x_j = \mathrm{cos} \Big( \frac{j\pi}{n} \Big), \ \ \ \ \ \ 0\leq j \leq n.  
\end{equation}
\bigskip

These points are the result of projecting equidistant points on the upper half of the unitary circle onto the real line.

\begin{figure}[H]
\centering
\includegraphics[scale=0.7]{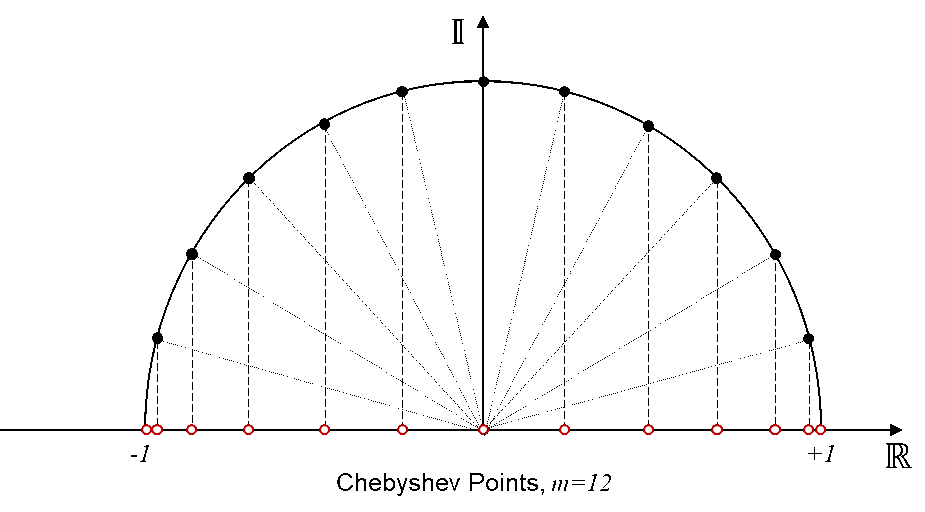}
\caption{Chebyshev points in one dimension.}
\end{figure}

The definition of Chebyshev points is given for an interval $[-1,1]$. This, however, can be extended to any interval $[a,b]$ by mapping $[-1,1]$ to $[a,b]$, with the aid of a linear transformation followed by a translation. Therefore, although most results in this Section are stated for functions defined on $[-1,1]$, or $[-1,1]^d$ in the case of $d$-dimensional functions, these are valid for more general domains $[a,b]$ and hyper-rectangles in higher-dimensions.
\bigskip

\begin{definition}
Let $n$ be a natural number. Let $v_0, \ldots, v_n$ be the values of a function $f$ on the Chebyshev points $x_0, \ldots, x_n$. The Chebyshev interpolant to these points, denoted by $p_n$, is the unique polynomial of order at most $n$ that interpolates the values $v_0, \ldots, v_n$ at the Chebyshev points $x_0, \ldots, x_n$.
\end{definition}
\bigskip

It is well known that given any $n+1$ points $x_0, \ldots, x_n$ and corresponding values $v_0, \ldots, v_n$ there is a unique polynomial of order at most $n$ that interpolates them. In particular, the Chebyshev interpolant is unique. For one-dimensional functions we have the following result which is one of the building blocks of the technique we propose in this paper. For a more detailed account and proofs, we refer to \cite{TrefethenTextbook}.
\bigskip

\begin{theorem}\label{thm: exponential convergence}
Let $f$ be an analytic function on the interval $[-1,1]$. Consider its analytical continuation to the open Bernstein ellipse $E_p$ of radius $\rho$, where it satisfies $|f(x)| \leq M$, for some $M$. Then for each $n\geq 0$
\begin{equation}
\|f-p_n \|_{\infty}\leq \frac{4M\rho^{-n}}{\rho - 1}  
\end{equation}

\noindent where $p_n$ is the Chebyshev interpolant to $f$ on the first $n+1$ Chebyshev points. 
\end{theorem}
\bigskip

\noindent Theorem \ref{thm: exponential convergence} says that very few interpolation points are needed to get a high degree of accuracy when the function is analytic. Within the context of pricing function approximation, given that most pricing functions are analytic (at least piecewise analytic), the use of Chebyshev interpolants makes sense.
\bigskip

In the context of numerical computing, it is important to have robust algorithms to run processes. It is often forgotten that the rounding errors on a computer can make compound with some algorithms making results unreliable. In the case of Chebyshev interpolants, the optimal way to evaluate them is with the Barycentric interpolation formula \cite{barycentricFormula}.
\bigskip

\begin{theorem}\label{thm: barycentric formula}
Let $x_0, \ldots, x_n$ be a grid of Chebyshev points and let $v_0, \ldots, v_n$ be values associated to this grid. Then the Chebyshev interpolant associated to these points is given by

\begin{equation}\label{eq: barycentric formula}
p(x) =  \sum\limits_{i=0}^{n}\textsc{\char13}\frac{(-1)^{i}v_i}{x-x_i} \Bigg/  \sum\limits_{i=0}^{n}\textsc{\char13}\frac{(-1)^{i}}{x-x_i} 
\end{equation}
\bigskip

\noindent for values of $x$ not on the grid. For the special case when $x = x_i$, then $p(x) = v_i$. The primes on the summation mean that when $i = 0$ or $i = n$, then the expression is multiplied by $1/2$.
\end{theorem}
\bigskip

\begin{remark}\label{rmk: barycentric props}
There are several advantages to using Equation \ref{eq: barycentric formula}. The first is that only the values of the function $f$ at Chebyshev points are needed in order to evaluate $p_n(x)$. The second is that evaluating such formula requires linear effort with respect to the degree of the polynomial. Thirdly, this formula is proven to be stable in floating point arithmetic for all $x$ within the domain of approximation \cite{BarycentricStability}. Also, it is scale-invariant, meaning that the formula does not change when we consider a general interval of the form $[a,b]$. The combination of Theorem \ref{thm: exponential convergence} and Theorem \ref{thm: barycentric formula} yield a technique that approximates pricing functions to a high degree of accuracy by calling it a small number of times, where the resulting approximator, a polynomial of low degree, can be evaluated in no time at all in a numerically stable manner (see \cite{TrefethenTextbook} for more details).
\end{remark}
\bigskip

\begin{remark}\label{rmk: barycentric speed one dim}
To give an idea of the speed of the barycentric interpolation formula within the context of pricing function approximation, a degree $9$ polynomial ($10$ Chebyshev points), which would give a high level of accuracy for most pricing functions in finance due to Theorem \ref{thm: exponential convergence}, takes around $100$ nanoseconds per evaluation on a standard computer  using a single core. If we are dealing with a risk calculation where $1,000$ evaluations need to be done, this would take $100$ microseconds or equivalently $0.0001$ seconds.
\end{remark}

\subsection{Multi-dimensional case}\label{sec: cheb interpolants multi dim}

Most pricing functions depend on a multitude of risk factors. For example, a Swaption depends on a multitude of interest rate tenors and implied volatilities. Therefore, to use Chebyshev interpolants within risk calculations we must be able to extend the notions presented so far to higher dimensions.
\bigskip

\begin{definition}\label{dfn: chebpts multi}
Let $A$ be a hyper-rectangle in $\mathbb{R}^n$. That is, $A$ is defined as the Cartesian product of one-dimensional closed and bounded intervals $I_i$, $A = I_1\times \cdots\times I_n$. Let $\chi_i$ be the Chebyshev points corresponding to the interval $I_i$, for all $i$, $1\leq i\leq n$. Let the number of Chebyshev points in $\chi_i$ be $m_i$. We define the mesh of Chebyshev points on $A$ generated by $\chi_1, \ldots, \chi_n$ as the Cartesian product of these sets, $\chi = \chi_1,\times\cdots\times \chi_n$. Note that the number of points on the multi-dimensional Chebyshev mesh is $m_1\cdots m_n$. 
\end{definition}
\bigskip

\begin{figure}
\centering
\includegraphics[width=10cm, height=11cm]{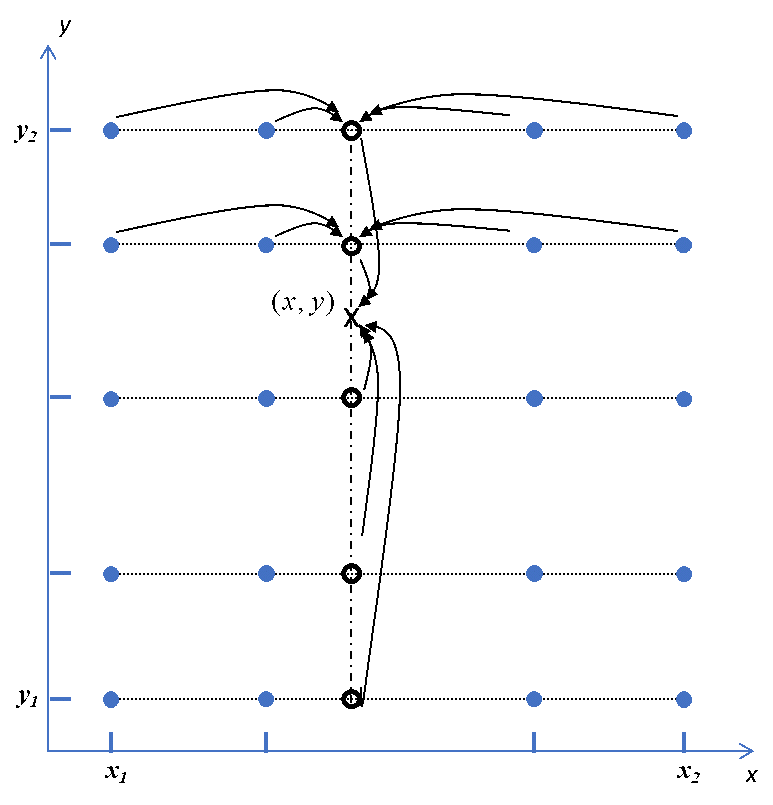}
\caption{Chebyshev mesh in two dimensions.}
\label{fig: 2d cheb mesh}
\end{figure}

As an illustration, Figure \ref{fig: 2d cheb mesh} shows a two-dimensional mesh. Say we have a $2$-dimensional function $f$ defined on $A$. Just as with the one-dimensional case, once the function has been evaluated on the mesh of Chebyshev points, a two-dimensional Chebyshev interpolant $p_{n,m}(x, y)$ is defined and is ready to be evaluated. 
\bigskip

There are a number of proposed approaches to evaluate multidimensional Chebyshev frameworks (see \cite{Trefethen3D}, \cite{Trefethen2D} and \cite{GlauParamOptPric}). We have found the following to be optimal within a practical setting. 
\bigskip

Without loss of generality, consider the point $(x, y)$. To evaluate the two-dimensional Chebyshev interpolant $p_{n,m}$ on $(x, y)$, consider the horizontal one-dimensional Chebyshev interpolants and evaluate them at $x$. This gives values on the black circles of Figure \ref{fig: 2d cheb mesh}. These black circles lie on the horizontal lines defined by the Chebyshev points on the $y$-axis. Hence, the black circles, along with the values on them obtained from the evaluation of the horizontal one-dimensional Chebyshev interpolants, define another one-dimensional Chebyshev interpolant (running vertically in Figure \ref{fig: 2d cheb mesh}) that can be evaluated on $y$. The result of the latter evaluation is the value of $p_{n,m}$ at $(x, y)$.
\bigskip

The evaluation just described can be extended in a straightforward manner to higher dimensions. If we start with a Chebyshev mesh of dimension $n$, we evaluate a collection of one-dimensional Chebyshev interpolants to reduce the problem from $n$ dimensions down to $n-1$ dimensions. Continuing this way, the problem is reduced to one dimension, where the evaluation of the resulting one-dimensional Chebyshev interpolant gives the result.
\bigskip

To put the time this takes to run into context, let us make an estimate based on the time taken for a $1$-dimensional Chebyshev interpolant (see Remark \ref{rmk: barycentric speed one dim}). Assume a $3$-dimensional Chebyshev interpolant. Moreover, assume $10$ Chebyshev points per dimension. This gives a total of $1000$ Chebyshev nodes on the whole mesh. Given the evaluation algorithm described above, the barycentric interpolation formula is called 111 times which gives, assuming $100$ nanoseconds per barycentric interpolation formula call, $10$ microseconds per $3$-dimensional Chebyshev interpolation evaluation. If there are $1000$ scenarios to evaluate in a risk calculation, this would roughly take 10 milliseconds or $0.01$ seconds for the whole calculation.
\bigskip

Just as in the case of dimension one, when the function $f$ is analytic, the convergence of Chebyshev interpolants is as good as can be expected. The following Theorem, recently published, can be found, along with its proof and details, in \cite{GlauParamOptPric}.
\bigskip

\begin{theorem}\label{thm: exponential multidim}
Let $f$ be a $d$-dimensional analytic function defined on $[-1,1]^d$. Consider its analytical continuation to a generalised Bernstein ellipse $E_p$, where it satisfies $\|f\|_{\infty}  \leq M$, for some $M$. Then, there exists a constant $C>0$, such that

\begin{equation}
\|f-p_n \|_{\infty}\leq C\rho^{-m} 
\end{equation} 

\noindent where $\rho=min_{(1\leq i\leq d)} \rho_i$, and $m=min_{(1\leq i\leq d)} m_i$. The collection of values $\rho_i$ define the radius of the generalised Bernstein ellipse $E_p$, and the values $m_i$ define the size of the Chebyshev mesh (see Definition \ref{dfn: chebpts multi}). For more details on Theorem \ref{thm: exponential multidim}, its proof and related results, see \cite{GlauParamOptPric}.
\end{theorem}
\bigskip

The combination of Theorem \ref{thm: exponential convergence}, Theorem \ref{thm: barycentric formula} and Theorem \ref{thm: exponential multidim} justifies the use of Chebyshev interpolants as a tool to approximate pricing functions. Leaving aside for now the curse of dimensionality that any mesh of points suffers from (that will be addressed in Section \ref{sec: orth sliding technique}), these results ensure that a small number of points on the mesh give high degrees of accuracy, while the resulting polynomials of low degree can be evaluated in a stable manner in virtually no time.
\bigskip

\begin{remark}
While a good proportion of pricing functions in finance are analytic, many have singular points and discontinuities. At both these types of points, the pricing function is not differentiable and hence not analytic. If the interpolants $p_n$ from Theorem \ref{thm: exponential multidim} are built on a domain that includes these points, there is no guarantee that they will converge exponentially. However, in the vast majority of cases, it is known where these singular points are; these points come, for example, in the form of payment dates, barriers and strikes. The domain of approximation can hence be split into sub-domains using these points as boundaries. Therefore, the restriction of the pricing function to each of these sub-domains is analytic. Chebyshev interpolants can be built on each of these sub-domains for which they converge to the pricing function, as described in Theorem \ref{thm: exponential multidim}.
\end{remark}
\bigskip

\section{The Orthogonal Chebyshev Sliding Technique}\label{sec: orth sliding technique}

Any technique that uses meshes of points suffers from the curse of dimensionality. That is, as the dimension of the mesh increases, the number of points on the mesh increases exponentially. As building Chebyshev interpolants requires evaluating the pricing function on such points, there will be dimensions for which building the interpolant is at least as expensive as evaluating the pricing function on all the scenarios required by the risk calculation.
\bigskip

Section \ref{sec: subsection sliding technique} deals with the Chebyshev sliding technique as a stand-alone method. In Section \ref{sec: subsection PCA} we show how the Chebyshev sliding technique can be combined with PCA to give the Orthogonal Chebyshev Slider, the technique used in the application presented in this paper.

\subsection{The Chebyshev Sliding Technique}\label{sec: subsection sliding technique}

A Chebyshev Slider consists of a collection of Chebyshev interpolants. The way these come together to constitute the slider is the following. Let $f$ be the function we are interested in approximating. Say $f$ is a function of dimension $n$. That is
\begin{equation}
f:A\longrightarrow \mathbb{R},
\end{equation}
\noindent for some hyper-rectangle $A$ in $\mathbb{R}^n$. Pick a point $z = (z_1,\ldots,z_n)$ in $A$ that we call the \emph{pivot point} for the slider. The function $f$ can be restricted to a lower dimensional space by fixing any of the dimensions with the values given by $z$ and letting the other ones vary. For example, let the first three dimensions vary and fix the rest with $z_4,\ldots,z_n$. Formally, this is equivalent to defining a new $3$-dimensional function $g_1$ which is the result of composing the injection
\begin{equation}
i:B\longhookrightarrow A,
\end{equation}
\noindent with $f$, where $B = \{(x_1,x_2,x_3) \in \mathbb{R}^3 \ | \ (x_1,x_2,x_3,z_4,\ldots,z_n)\in A\}$, and $i$ takes $(x_1,x_2,x_3)$ to $(x_1,x_2,x_3,z_4, \ldots,z_n )$, yielding
\begin{equation}
g_1:B\longhookrightarrow  A\longrightarrow \mathbb{R}.
\end{equation}

Clearly, a Chebyshev interpolant can be built for $g_1$, and since its dimension is $3$, the curse of dimensionality will not affect the building process (i.e. we are only evaluating the function $g_1$ on a $3$-dimensional Chebyshev mesh). This Chebyshev interpolant, denoted by $s_1$, is called a \emph{Chebyshev Slide}. A \emph{Chebyshev Slider} is a collection of slides, say $\{s_1,s_2,\ldots,s_k\}$, where each of the variables in the domain of $f$ is part of the domain of one and only one slide. In this way, the sum of the dimensions of the slides $dim(s_1 )+\cdots +dim(s_k )$ is $n$.
\bigskip

As the Chebyshev Slider consists of a collection of slides, which are Chebyshev interpolants, the building of the slider consists of the building of at least one Chebyshev interpolant ($k$ in our example). If there are $k$ slides and each has $l_i$ Chebyshev points on which the pricing function is evaluated, the total number of evaluations needed to build the slider is $l_1+\ldots+l_k$. If the slides are typically of $1$, $2$ or $3$ dimensions, for which at most a few tens of points are needed, the total number of points needed for the slider is substantially smaller than for an $n$-dimensional mesh, thus side-stepping the curse of dimensionality.
\bigskip

When building a Chebyshev Slider, one must choose two things. First, the number of slides. Second, the number of exploration points of each slide. These two choices define the configuration of the slider. Changing the configuration will change the accuracy of approximation and the effort to build it. The smaller the number of slides, the higher the accuracy but the greater the dimension of the meshes built, which translates into higher building costs. The greater the number of slides, the poorer the approximation but the greater the computational gains.
\bigskip

The evaluation of the slider is done in the following way. Let $x=(x_1,\ldots,x_n)$ be a point in the domain of $f$ and let $v$ be the value of $f$ at the pivot point $z$, $v=f(z)$. For ease of presentation, assume without loss of generality that $dim(s_1 )=3,dim(s_2 )=2$, and $dim(s_i )=1$, for $3\leq i\leq k$. Then the value of $f$ at $x$ is given by
\begin{equation}\label{eq: slider eval}
f(x)=v + s_1 (x_1,x_2,x_3 ) - v +  s_2 (x_4,x_5 ) - v + S_3 (x_6 ) - v+\cdots + s_k (x_n ) - v.
\end{equation}

From now on, sliders will be represented by the $k$-tuple obtained from the dimensions of its constituent sliders. For the example above, where the slider consists of the slides $\{s_1,s_2,\ldots ,s_k\}$, we use the $k$-tuple $\{dim(s_1), dim(s_2),\ldots,dim(s_k)\}$. 
\bigskip

\begin{remark}
Equation \ref{eq: slider eval} should remind the reader of Taylor approximation. Indeed, Chebyshev interpolants are to the slider what partial derivatives are to Taylor approximation. However, instead of having partial derivatives of order one, or at most two, which often fail to fully capture the function in the domain spanned by the variables concerned, the Chebyshev interpolants, or slides,  $s_1,\ldots,s_k$, approximate their corresponding restrictions of $f,g_1,\ldots,g_k$, to a very high degree of accuracy. Given that the evaluation of the Chebyshev Slider relies on the evaluation of the Chebyshev interpolants that constitute it, it is numerically stable and extremely efficient. 
\end{remark}
\bigskip

\begin{remark}\label{rmk:barycentric speed}
To get an idea of the efficiency, consider an example of a $20$-dimensional Chebyshev Slider and assume a slider configuration where the first slider has dimension $3$, while the remaining have dimension one. Based on the estimates from Remark \ref{rmk: barycentric props}, the time it takes to evaluate a  slider of this type would be dominated by the slider of dimension $3$ giving a rough estimate of $10$ microseconds per evaluation. For a risk calculation with $3000$ evaluations, which corresponds to finding the period of stress within FRTB IMA, the whole run would take only $0.03$ seconds.
\end{remark}

\subsection{Orthogonal dimensionality reduction}\label{sec: subsection PCA}

In some applications, such as the one we deal with in this paper, reducing the dimensionality of the risk factor space, before applying the Chebyshev Sliding Technique, improves the computational savings. Sometimes the dimensionality of the pricing functions to be approximated with Chebyshev Sliders is too great. For example, Swaptions depend on a few hundredths of dimensions. Even if the slides in the slider have low dimensionality, the number of points needed to build the slider can be substantial. If a dimensionality reduction technique can reduce the dimension of the risk factor space without losing much information, then the slider can be built on a space with much smaller dimension, substantially reducing the building cost.
\bigskip

The technique chosen to reduce the dimensionality of the space of risk factors in the current application is Principal Component Analysis (PCA). It is a technique which is fast and easy to use. It does not have the complexity introduced by more advanced Machine Learning algorithms which require hyper-parameter optimisation. It requires little monitoring. Moreover, it works well on financial risk factors, as many of these are highly correlated. For example, an interest rate swap depends on several tens of interest rates. All these are highly correlated hence PCA can reduce the dimensionality of the set to only a few factors without losing much information.
\bigskip

PCA reduces the dimension of the space of risk factors by projecting the data points onto a hyper-plane that minimises the average of the projection distances of the data. The reduction in dimensionality is then achieved by a change of coordinates. This change of coordinates is precisely the transformation that lets us express a point in high dimensions as a point in a lower dimensional space. As changes in coordinates are given by linear isomorphisms, it has an inverse, which we denote by $T^{-1}$
\begin{equation}
T^{-1}: \mathbb{R}^k\longrightarrow\mathbb{R}^n.
\end{equation}
This transformation changes the low dimensional expression of any given point $x$ to a high dimensional expression $(k\leq n)$. By composing $T^{-1}$ with the pricing function $f$, we obtain a function 
\begin{equation}
g: \mathbb{R}^k\longrightarrow\mathbb{R}^n\longrightarrow\mathbb{R}.
\end{equation}
\noindent for which we build a Chebyshev Slider.
\bigskip

For example, if $f$ is the pricing function of a Swaption, $n$ is a few hundreds. However, by applying PCA to the set of rates and the set of volatilities, we can reduce the dimensionality, as can be seen in Section \ref{sec: swaption results}, to around $20$ without losing much information. The resulting function $g$, for which a slider is built, is a $20$-dimensional function for which building a slider is very cheap.
\bigskip

It is important to note that building a Chebyshev Slider of high dimensionality (say, $30$ or $50$) is perfectly possible. This lets us keep a large number of Principal Components, keeping the loss of accuracy coming from PCA to a minimum.
\bigskip

\begin{remark}
It must be noted, however, that in principle, all that is needed is a dimensionality reduction technique that comes equipped with a function that can change the transformed set of data points (the ones with low dimension) to data points with the original dimensionality. This function would play the role of $T^{-1}$ above. At the time of writing this paper, the authors are exploring the use of non-linear transformations via autoencoders.
\end{remark}
\bigskip

When a dimensionality reduction technique is used in tandem with Chebyshev Sliders, there are parameters, other than the ones coming from the slider, that need to be specified. In the present case study, where PCA is used, the dimension we reduce the risk factors down to becomes another parameter to consider. As we present in Section \ref{sec: Results}, a collection of Chebyshev Sliders and PCA reductions (that we will call Orthogonal Chebyshev Sliders) were tested to see the impact of different configurations on the results.

\section{Proof of Concept description}\label{sec: PoC description}

\subsection{Portfolios and success criteria}

In this Section we describe the case study carried out within the risk systems of a tier-one bank.\footnote{Permission to make public the results was given.} There were two portfolios used: one with $635$ Interest Rate Swaps; the second with $425$ Vanilla Swaptions. Both dependent on a single currency. For the portfolio of Swaps it was agreed to compute its price on $3,131$ historic shocks. In the case of the portfolio with Swaptions, $3,108$ historic shocks.\footnote{In both cases, the number of shocks corresponds to the available data for a 10-year period which includes a period of stress.} In the latter case, the shocks were priced twice, as stipulated by the IMA-FRTB capital calculation framework: once for the $10$-day liquidity horizon, which corresponds to all risk factors being shocked; the second for the $60$-day liquidity horizon, which corresponds to rates being constant and volatilities shocked. From each of the three price distributions a P$\&$L was obtained and from each P$\&$L distribution an ES value computed.
\bigskip

The main objective of the agreed scope of the Proof of Concept (PoC) was to compare the brute force approach and the Orthogonal Chebyshev sliding technique in terms of speed and accuracy  (accuracy at the level of ES).\footnote{The brute force approach being the one where the Front Office pricing functions are called on every historic shock.} The idea being to benchmark the Orthogonal Chebyshev sliding technique with the most accurate albeit expensive methodology.
\bigskip

The calculation of prices on a distribution of historic shocks that correspond to the last ten years is, computationally speaking, the bulk of what is needed for the calculation of the period of stress. The latter was not computed in this PoC as that was not part of the scope, but high pricing accuracy on $10$ years of historic shocks, which was obtained  (see Section \ref{sec: Results}), is the key to accurately computing the period of stress with any alternative technique.
\bigskip

A further advantage of building Orthogonal Chebyshev Sliders for $10$ years of data is that they can also be used to price the most recent $250$ scenarios needed for the capital calculation within FRTB IMA. As mentioned, this exercise was not part of the agreed scope of the PoC. However, to illustrate this use case, the ES value on today's $250$ shocks is shown in Section \ref{sec: Results} for each of the portfolios.
\bigskip

There were three success criteria in the PoC.

\begin{enumerate}
	\item To build Orthogonal Chebyshev Sliders with $50\%$ less calls to the pricing function than if the calculation had been done in a brute force way. This represents a computation reduction, by using the Orthogonal Chebyshev sliding technique, of at least $50\%$.  To see why, it is important to bear in mind that evaluating Orthogonal Chebyshev Sliders amounts to the evaluation of a collection of polynomials with the barycentric interpolation formula (see Theorem \ref{thm: barycentric formula}). Therefore, the time it takes to evaluate the risk scenarios with Orthogonal Chebyshev Sliders is negligible (see Remark \ref{rmk:barycentric speed}) compared to the time it takes to build them when we call the Front Office pricing functions.
	\item The difference between the ES values calculated using the brute force approach and using Orthogonal Chebyshev Sliders should not differ in no more than $10\%$. 
	\item Third, it was important to show that the Orthogonal Chebyshev Slider technique could be implemented within the bank's systems in a seamless manner.
\end{enumerate}
\bigskip

With the interest of understanding the effects that changing the parameters of the PCA and Orthogonal Chebyshev Slider have on the accuracy and speed of the technique, several different parameters configurations were tested. This is crucial to understand the stability of the methodology as ideally one would want to change these parameters as little as possible from day to day. The result of these are presented in Section \ref{sec: Results}.
\bigskip

The accuracy of the Orthogonal Chebyshev Slider technique was ultimately measured on the ES values obtained. However, the P$\&$L distributions obtained by the slider were also tested. This was done in two ways. First, by computing the correlation of the P$\&$L  obtained using the brute force approach with the P$\&$L  obtained using Orthogonal Chebyshev Sliders. Second, by running the Kolmogorov-Smirnov test on these same two P$\&$Ls. The idea behind these tests was to rule out the existence of any statistical evidence concerning the differences between the benchmark distribution (brute force) and the one obtained with the Orthogonal Chebyshev Slider technique.

\subsection{Different liquidity horizons}\label{sec: diff liq horizons}

There is another point which is relevant in the calculation of capital within FRTB IMA and of importance to the Orthogonal Chebyshev Sliding Technique. For each trade, there are as many ES values that need to be computed as liquidity horizons. In the present taste case, the portfolio of Vanilla Swaptions was exposed to two liquidity horizons: the $10$-day one where all risk factors are shocked and the $60$-day one where the volatilities are shocked while rates remain constant. Another objective of the PoC was to build only one Orthogonal Chebyshev Slider for each trade irrespective of the liquidity horizons. Specifically, the Orthogonal Chebyshev Slider built for the $10$-day liquidity horizon ES calculation would be used for the $60$-day ES calculation. In this way, the technique would incur in the computational burden from building the slider only once.
\bigskip

The way this was achieved was by using multiple PCA transformations. More specifically, risk factors with different liquidity horizon were transformed by different PCA models. For example, in the case of the Vanilla Swaptions, which depend on interest rates (shocked only in the $10$-day liquidity horizon) and implied volatilities (shocked in both $10$-day and $60$-day liquidity horizons), a PCA function, denoted by $p_1$, would be trained on the shocked interest rates, and a second one, independent of the first, denoted by $p_2$, trained on the shocked implied volatilities. An Orthogonal Chebyshev Slider would then be built (as described in Section \ref{sec: subsection PCA}) on the domain obtained by putting together the projections of these PCA models. 
\bigskip

For the $60$-day liquidity horizon ES calculation, the volatility shocks are the same ones $p_2$ got trained on (i.e. Shocked implied volatilities), while the rates are fixed to today's shock.\footnote{Usually the the zero shock.} To use the Orthogonal Chebyshev Slider built for the $10$-day liquidity horizon in the computation of the $60$-day liquidity horizon, the $60$-day liquidity horizon shocks need to be projected, using the PCA models trained for the $10$-day case, within the domain of the Orthogonal Chebyshev Slider. Moreover, for this approach to be of any use, the error introduced by this projection should not be material. This is indeed the case since the volatilities are projected through $p_2$ while today's shock is projected under $p_1$. The former is the same as what had been done in the $10$-day liquidity horizon case. For the latter, as today's shock belongs to the collection of shocks $p_1$ got trained with, the error of projecting today's shock under $p_1$ will not be great and the projection will be within the domain of the Orthogonal Chebyshev Slider. This ensures the Orthogonal Chebyshev Slider built for the $10$-day liquidity horizon case can be used once again for the $60$-day liquidity one. 
\bigskip

Notice this approach works regardless of the number of liquidity horizons. The key is to use at least as many PCA models as collections of risk factors with individual liquidity horizons and train them on the shocks that correspond to the $10$-day liquidity horizon. The consequence is that the only Orthogonal Chebyshev Slider that needs to be built is the one corresponding to the $10$-day liquidity horizon by-passing the heavy part of the process for any other liquidity horizon.
\bigskip

\section{Results}\label{sec: Results}

In this Section we present and discuss the results for the portfolios of Swaps and Swaptions.
\bigskip

For each dimensionality reduction, three different Orthogonal Chebyshev Slider configurations were tested. The same  configurations were chosen for Swaps and Swaptions. Once a dimension has been fixed, two parameters need to be specified to define a slider: the dimension of each of the slides or Chebyshev interpolants, and the number of interpolation points for each slide. 
\bigskip

For simple products, such a Swaps, a few points (e.g. $5$) will approximate the portion of the function approximated by the slide to a high degree of accuracy. In the case of Swaptions, or any other product with curvature, only a couple of points more are needed, if at all, due to the exponential convergence of Chebyshev interpolants.\footnote{Remember that sliders are collections of Chebyshev interpolants.} Any more points will only incur in greater building times when no greater accuracy is needed. For simplicity, the number of points for each dimension in each slide was fixed to $5$, for both Swaps and Swaptions.
\bigskip

The dimension of the slides was chosen to be the simplest possible. We started with the most basic slider and the one that gives the least number of interpolation points; that is, the slider with configuration $\{1,1,\ldots,1\}$.\footnote{ For a description of this nomenclature we refer to Section \ref{sec: subsection sliding technique}.}  This slider only captures linear combinations of the one-dimensional cross-sections of the function along the PCA components. To capture more complex movements of the function, one must define sliders with slides that have dimensions greater than one. In this case it is important to consider the balance between information lost (sliders with slides of low dimension incur in greater information lost) and computational gain (sliders with slides of low dimension give greater computational gains). Given that PCA components capture the directions of greater variance in the data, it was decided to group the most important components together and build a single slide for them. Therefore, apart from the configuration $\{1,1,\ldots,1\}$, two others were considered: $\{2,1,\ldots,1\}$ and $\{3,1,\ldots,1\}$.
\bigskip

For each of the slider configurations just mentioned, a range of PCA dimensions was considered. The way in which such range was chosen is the following. The Swaps considered in the PoC depended on two zero rates curves. The values of zero rate curves exhibit high correlation. It is well known that only a few PCA components capture a good degree of the variance, therefore, the range of PCA dimensions chosen was $3, 5, 10$ and $20$.
\bigskip

The Swaptions used in this PoC depend on the same zero rate curves as Swaps plus a surface of implied volatilities. As explained in Section \ref{sec: diff liq horizons}, two PCA transformations were used with Swaptions so that the Orthogonal Chebyshev Slider built for the $10$-day liquidity horizon movements could be re-used for the $60$-day liquidity horizon. Therefore, the dimensionality reduction is the sum of the dimensionality reduction for zero rates and the dimensionality reduction for volatilities. For simplicity the same dimension was chosen for zero rates and for volatilities. The dimensions considered were $10, 20, 30$ and, $50$.\footnote{Which corresponds to dimensions $5, 10, 15, 25$ for zero rates and for volatilities.} In the case of Swaptions, we do not expect dimensions less than $10$ to work; dimension $50$ or greater will probably not give any accuracy improvements. 
\bigskip

It should be noted, however, that the dimension range for which good results are obtained depends on the quality of the data and the number of types of risk factors. Part of the exercise within the PoC was to identify this range and see how sensitive the accuracy of the methodology is to a change in PCA dimensionality reduction. This sensitivity is presented in Figure \ref{fig: plot sens swaps} for Swaps, and Figure \ref{fig: plot sens swaptions 10d} and Figure \ref{fig: plot sens swaptions 60d} for Swaptions.
\bigskip

There are two sources of error when using the Orthogonal Chebyshev Slider technique. The first is due to the PCA dimensionality reduction. The second is due to the approximation of the Orthogonal Chebyshev Slider on the projected set of risk factors. This is the reason why we present the results as shown, for example, in Figure \ref{fig: plot 3000 swaps}. This figure corresponds to the results for Swaps using the slider $\{1,\ldots, 1\}$. The corresponding figures for sliders $\{2,1,\ldots,1\}$ and $\{3,1,\ldots,1\}$ can be found in Appendix \ref{sec: Appendix diff configs results}. The corresponding figures for Swaptions can be found in Section \ref{sec: swaption results} and Appendix (\ref{sec: Appendix diff configs results}).
\bigskip

Each of these figures consists of four plots, one for each dimensionality reduction. Each of these plots consists of four graphs. The first three graphs are scatter plots, the fourth is a bar plot. The first scatter plot is a comparison between the P$\&$Ls obtained through brute force for the whole portfolio of Swaps on the $3,131$ scenarios ($x$-axis) and the P$\&$Ls obtained by evaluating the pricing function on the scenarios projected under the PCA transformation ($y$-axis). The idea of this scatter plot is to give visual evidence of the error at the level of P$\&$Ls that comes from the PCA dimensionality reduction. The second diagonal scatter plot makes a comparison of the P$\&$Ls obtained by evaluating the pricing function on the projected scenarios ($x$-axis) and the P$\&$Ls obtained by the Chebyshev Sliding Technique ($y$-axis). This gives visual evidence of the error incurred by the Orthogonal Chebyshev Slider alone. The third is a comparison of the P$\&$Ls obtained through brute force versus the P$\&$Ls obtained by the Orthogonal Chebyshev Slider. This shows the resulting accuracy, when we combine PCA with Chebyshev Sliders.
\bigskip

The idea behind plotting all three scatter plots is to identify where errors are coming from. If there is significant error between the P$\&$Ls obtained from the Orthogonal Chebyshev Slider and the brute force P$\&$Ls, we want to see whether the error is due to the Chebyshev Slider or the PCA projection. This can give an indication that perhaps the dimensionality of the PCA should be increased, for example.
\bigskip

The last plot in each figure is a bar plot. This shows the ES value obtained via brute force (blue bar) compared to the value of the ES obtained via the Orthogonal Chebyshev Slider (red dot). A $10\%$ error bar has also been added around the brute force ES as reference, as it was the agreed accuracy threshold for the PoC.
\bigskip

Each figure, like the one described above, comes with a table that contains the following numerical values: the ES relative error, percentage of computational savings, P$\&$L correlations and $p$-values for the Kolmogorov-Smirnoff tests. The P$\&$L correlations and $p$-values from the Kolmogorov-Smirnoff tests were added to assess the overall quality of the P$\&$L approximations obtained from the combination of PCA and Orthogonal Chebyshev Sliders. In particular, the Kolmogorov-Smirnoff test measures whether there is evidence that the two P$\&$L distributions being compared (brute force versus Orthogonal Chebyshev Slider) are not the same.\footnote{Low $p$-values – for example, lower than $0.05$ – give an indication that the two distributions being compared are not the same.}

\subsection{Swaps}\label{sec: swap results}

\begin{figure}[tp]
\centering
\includegraphics[width=14cm, height=12cm]{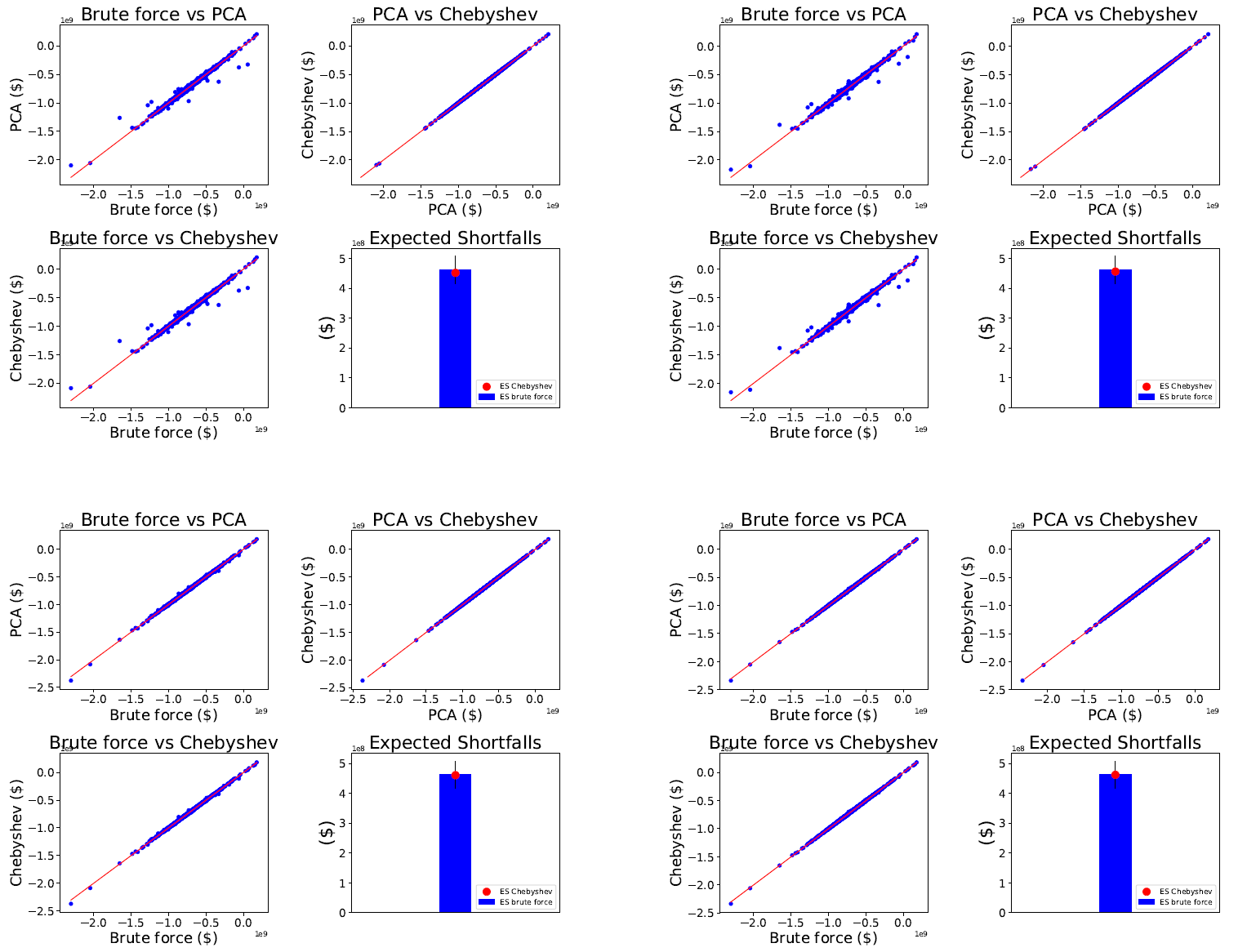}
\caption{Portfolio of Swaps, slider configuration $\{1,1,\dots,1\}$. Top left: PCA dim $3$. Top right: PCA dim $5$. Bottom left: PCA dim $10$. Bottom right: PCA dim $20$.}
\label{fig: plot 3000 swaps}
\end{figure}

From Figure \ref{fig: plot 3000 swaps} it is easy to see that even for very low dimensionality reductions the results obtained are outstanding. Its corresponding Table \ref{tab: swaps} shows that with dimension $3$ the ES relative error is already of $2.20\%$. As the PCA dimension increases there is an improvement in the ES error and the P$\&$L correlation. Meanwhile, at no point does the KS $p$-value show any evidence that the P$\&$L distributions are different. The increase in dimension comes with a detriment in computational savings being $96.77\%$ at its worse. In any case, both the worse accuracy, which is of $2.20\%$, and the worse computational gain, which is of $96.77\%$, are outstanding.
\bigskip

\begin{figure}[t]
\centering
\includegraphics[scale=0.45]{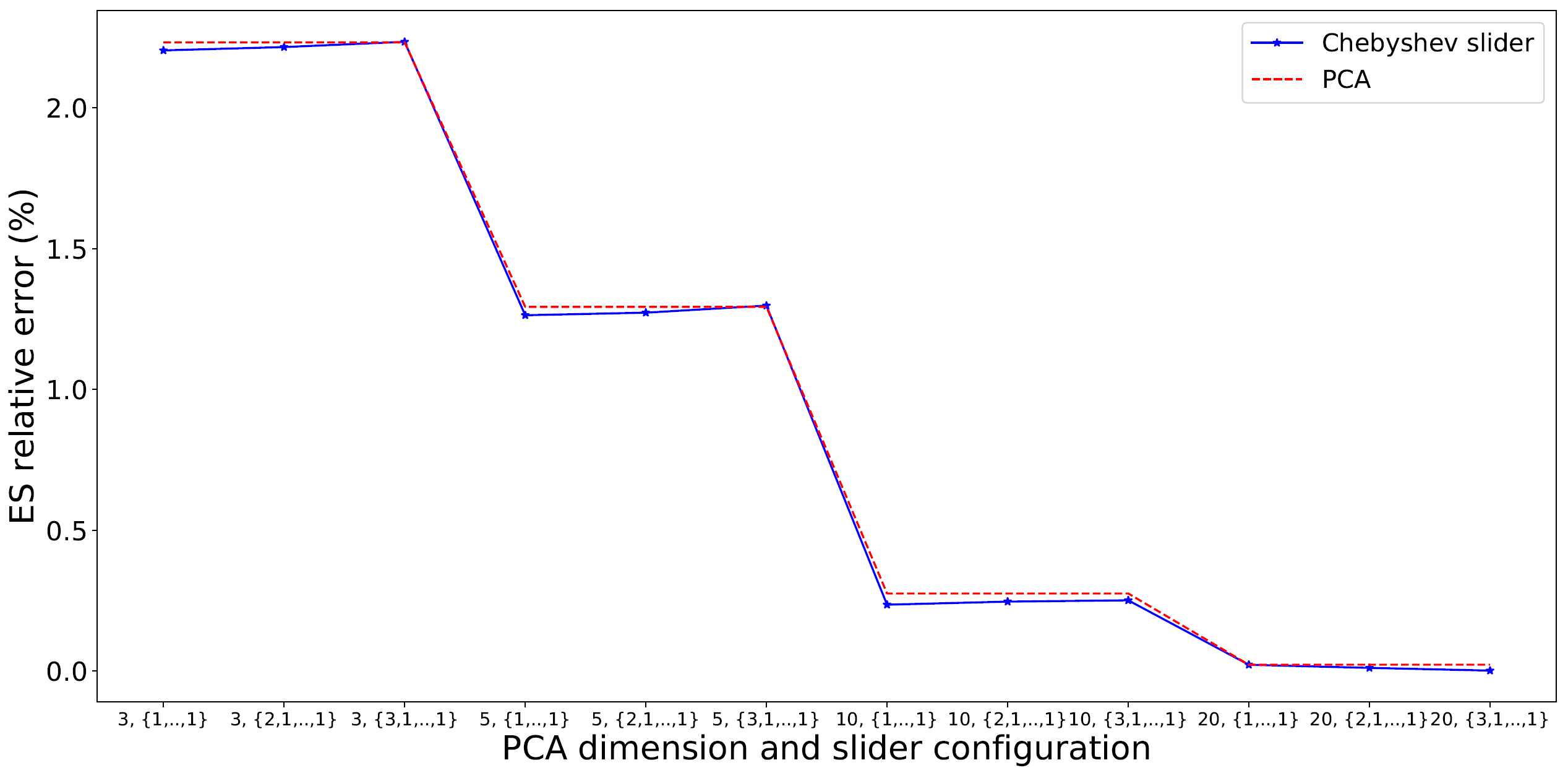}
\caption{This figure shows how the Orthogonal Chebyshev Slider relative ES error changes as dimensionality and slider configuration change for Swaps.}
\label{fig: plot sens swaps}
\end{figure}

\begin{figure}[H]
\centering

\includegraphics[width=14cm, height=12cm]{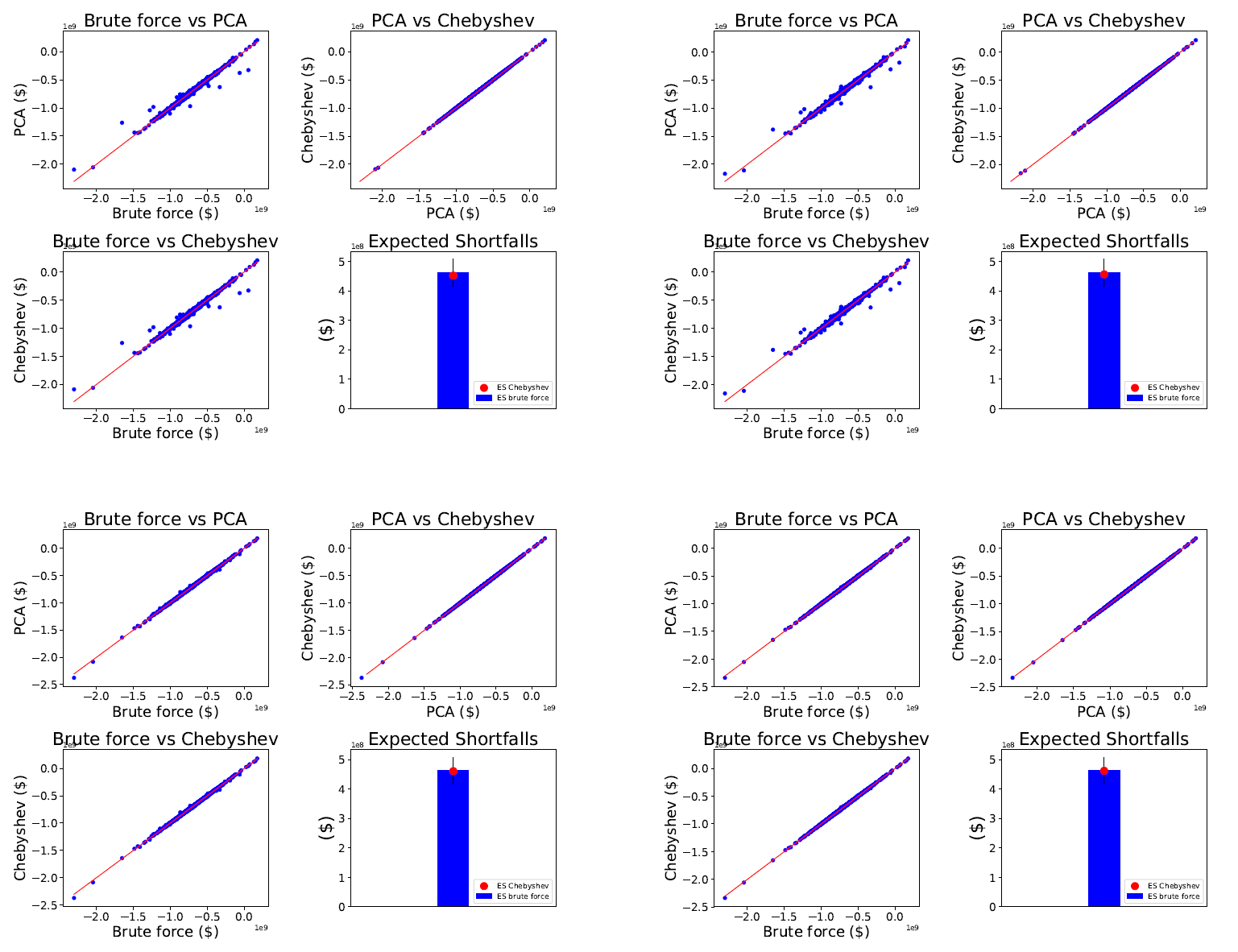}
\caption{Portfolio of Swaps, slider configuration $\{2,1,\dots,1\}$. Top left: PCA dim $3$. Top right: PCA dim $5$. Bottom left: PCA dim $10$. Bottom right: PCA dim $20$.}
\label{fig: plot 3000 swaps 2s}
\end{figure}

\begin{figure}[H]
\centering
\includegraphics[width=14cm, height=12cm]{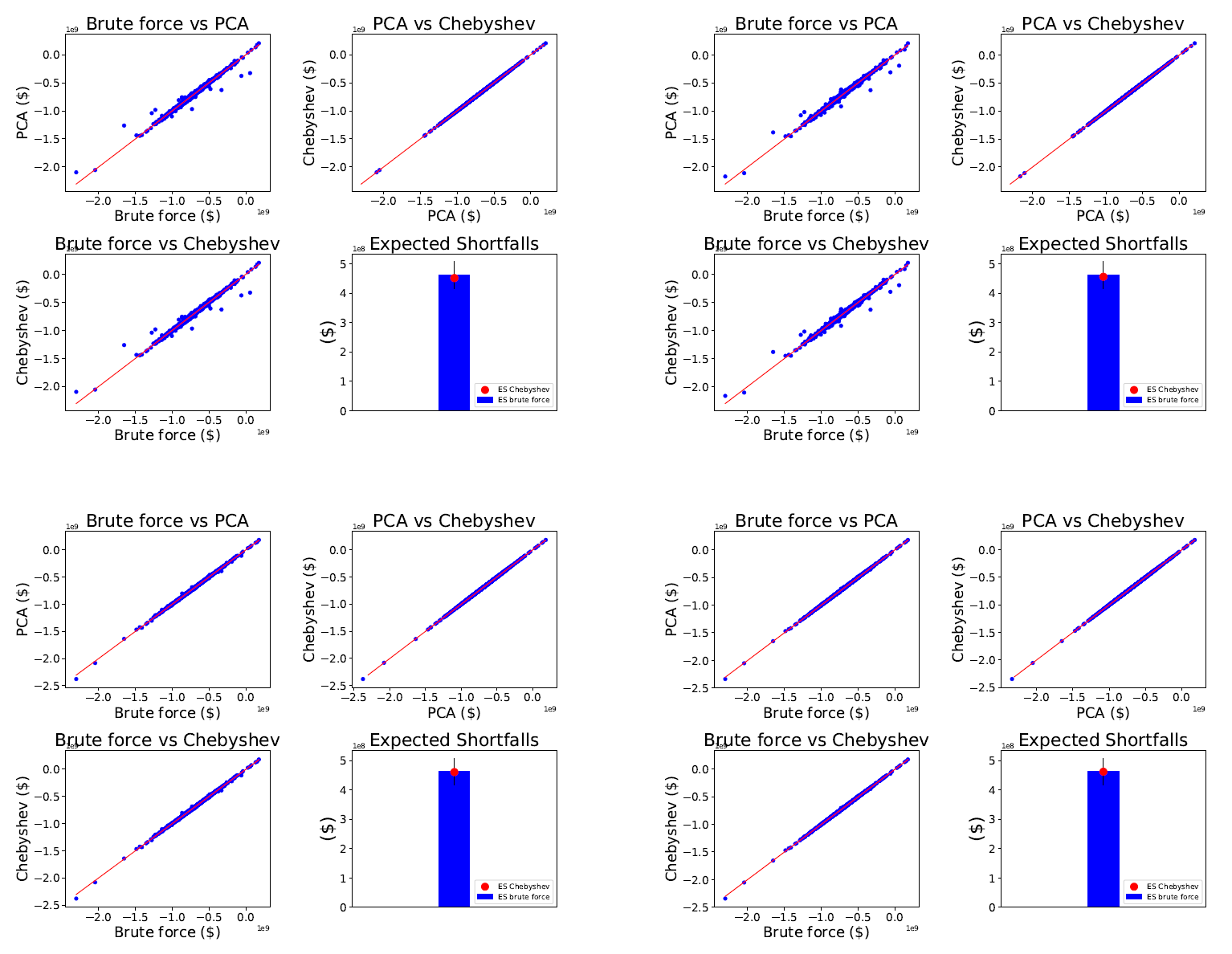}
\caption{Portfolio of Swaps, slider configuration $\{3,1,\dots,1\}$. Top left: PCA dim $3$. Top right: PCA dim $5$. Bottom left: PCA dim $10$. Bottom right: PCA dim $20$.}
\label{fig: plot 3000 swaps 3s}
\end{figure}

As can be seen from Figure \ref{fig: plot 3000 swaps 2s} and Figure \ref{fig: plot 3000 swaps 3s}, along with their corresponding Table \ref{tab: swaps 2s} and Table \ref{tab: swaps 3s}, changing the configuration to $\{2,1,\ldots,1\}$ and $\{3,1,\ldots,1\}$ did not impact the quality of the results. This can be much more clearly seen in Figure \ref{fig: plot sens swaps} where the Chebyshev configurations are plotted versus ES errors. The biggest improvement in ES error is obtained in every increase of PCA dimensionality reduction and not when the configuration of the slider is changed. For Swaps, clearly the best option is to keep the simplest and least expensive configuration which is $\{1,1,\ldots,1\}$. This is to be expected as Swaps exhibit linear behaviour.
\bigskip

\begin{figure}
\centering
\includegraphics[scale=0.60]{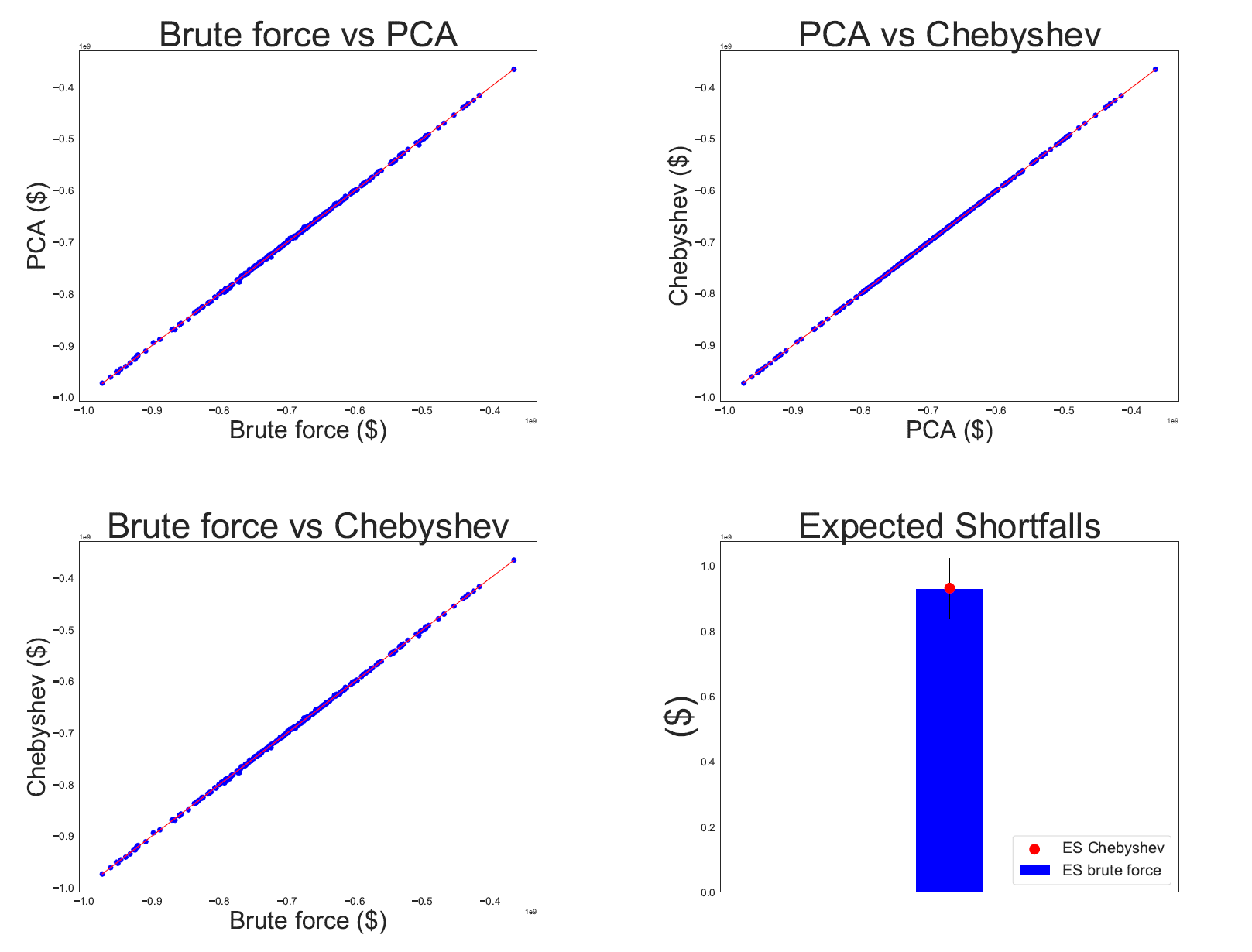}
\caption{Portfolio of Swaps, slider configuration $\{1,1,\ldots,1\}$, PCA dimension $10$, evaluated on the most recent $250$ scenarios.}
\label{fig: plot 250 swaps}
\end{figure}

The results presented in this Section were obtained for $3,131$ scenarios. Using the same Orthogonal Chebyshev Slider built for these many scenarios but evaluating them only on the most recent $250$, give the results shown in Figure \ref{fig: plot 250 swaps}. The slider chosen to compute the last $250$ scenarios has dimension $10$ and configuration $\{1,\ldots,1\}$. We expect any of the four to have yielded similar results, but dimension $10$ gives accuracy values of less than $1\%$.
\bigskip

\begin{table}[H]
\centering
\begin{tabular}{|l|l|l|l|l|}
\hline
Slider $\{1, \ldots, 1\}$ & PCA dim $3$ & PCA dim $5$ & PCA dim $10$ & PCA dim $20$\\
\hline
ES relative error & $2.20\%$ & $1.26\%$	& $0.24\%$ & $0.02\%$\\
Computational savings & $99.49\%$ & $99.17\%$ & $98.37\%$ & $96.77\%$\\
Correlation & $0.99$ & $1.00$ & $1.00$ & $1.00$\\
KS $p$-value & $0.89$ & $0.94$ & $1.00$ & $1.00$\\
\hline
\end{tabular}
\caption{Portfolio of Swaps, slider configuration $\{1,1,\ldots,1\}$.}
\label{tab: swaps}
\end{table}

Based on the number of times the pricing function was called to build such slider, the computational gain stands at $80\%$ if we only consider $250$ scenarios to evaluate. However, this slider was built for $3,131$ scenarios. If only $250$ scenarios need to be evaluated (as opposed to also $3,131$) a lower PCA dimension (i.e. higher computational savings) can be used with equally accurate results as for dimension $10$. Reducing the PCA dimension would give less Chebyshev points to evaluate under the pricing function, increasing computational reduction gains. For example, it is likely that $5$ dimensions or less would suffice, giving computational gains of over $90\%$. We refer to Remark \ref{rmk: re-usability} for a more detailed account of the implications of using Orthogonal Chebyshev Sliders for the daily computation of ES.
\bigskip

\FloatBarrier

\subsection{Swaptions}\label{sec: swaption results}

\subsubsection{10-day liquidity horizon}\label{sec: 10-day liq horiz}

\begin{figure}[tp]
\centering
\includegraphics[width=14cm, height=12cm]{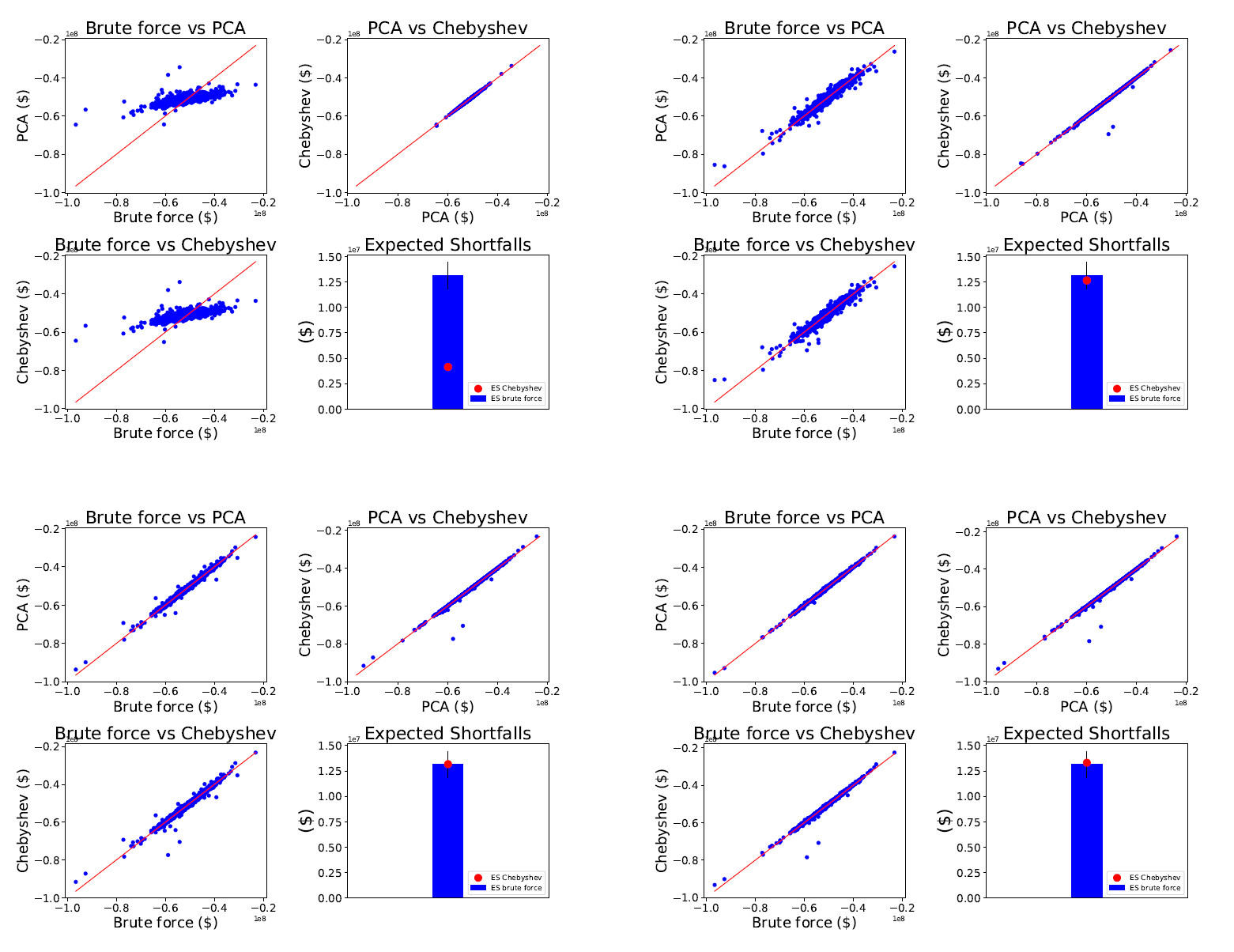}
\caption{Portfolio of Swaptions, slider configuration $\{1,1,\dots,1\}$, on $10$-day liquidity horizon. Top left: PCA dim $10$. Top right: PCA dim $20$. Bottom left: PCA dim $30$. Bottom right: PCA dim $50$.}
\label{fig: plot 3000 swaption 10d}
\end{figure}

For Swaptions, the Orthogonal Chebyshev Slider configurations used were the same as for Swaps. However, as Swaptions have many more risk factors, especially in the $10$-day liquidity horizon, the range of PCA dimensions considered was higher: $10, 20, 30$ and $50$. As can be seen from the top left-hand plot in Figure \ref{fig: plot 3000 swaption 10d}, dimension $10$ does not capture the P$\&$L brute force distribution very well. Moreover, we can see the problem is dimensionality of the PCA as the PCA versus Chebyshev plot shows very high accuracy. As soon as we increase the dimension to $20$ results improve considerably. With dimension $20$, regardless of the configuration, we already have an ES error of less than $4\%$. As soon as we reach dimensions $30$ and above, the accuracy drops below $1\%$.
\bigskip

When it comes to computational savings, the higher the PCA dimension the lower the gains. This makes sense as domains of approximation with higher dimensions require interpolation grids with more points.
\bigskip

The computational savings are once again staggering for the case tested. Even in the extreme case of $50$ dimensions, the savings are of $91.9\%$ for the slider $\{1,\ldots,1\}$. Given that ES accuracy is well under $1\%$ for dimension $30$, one could choose this dimension hence increasing the computational gain to $95.14\%$. If the accuracy appetite is fine with the range $1\%-5\%$, a lower dimensionality, such as $20$, can be chosen, increasing the computational gain further to $96.75\%$.

\begin{table}[p]
\centering
\begin{tabular}{|l|l|l|l|l|}
\hline
Slider $\{1, \ldots, 1\}$ & PCA dim $10$ & PCA dim $20$ & PCA dim $30$ & PCA dim $50$\\
\hline
ES relative error & $68.47\%$ & $3.75\%$ & $0.36\%$	& $1.41\%$\\
Computational savings & $98.36\%$ & $96.75\%$ & $95.14\%$ & $91.92\%$\\
Correlation & $0.75$ & $0.97$ & $0.99$ & $0.99$\\
KS $p$-value & $0.00$ & $0.93$ & $0.83$ & $1.00$\\
\hline
\end{tabular}
\caption{Portfolio of Swaptions, slider configuration $\{1,1,\ldots,1\}$ on $10$-day liquidity horizon.}
\label{tab: swaptions 10d}
\end{table}

\begin{figure}[t]
\centering
\includegraphics[scale=0.45]{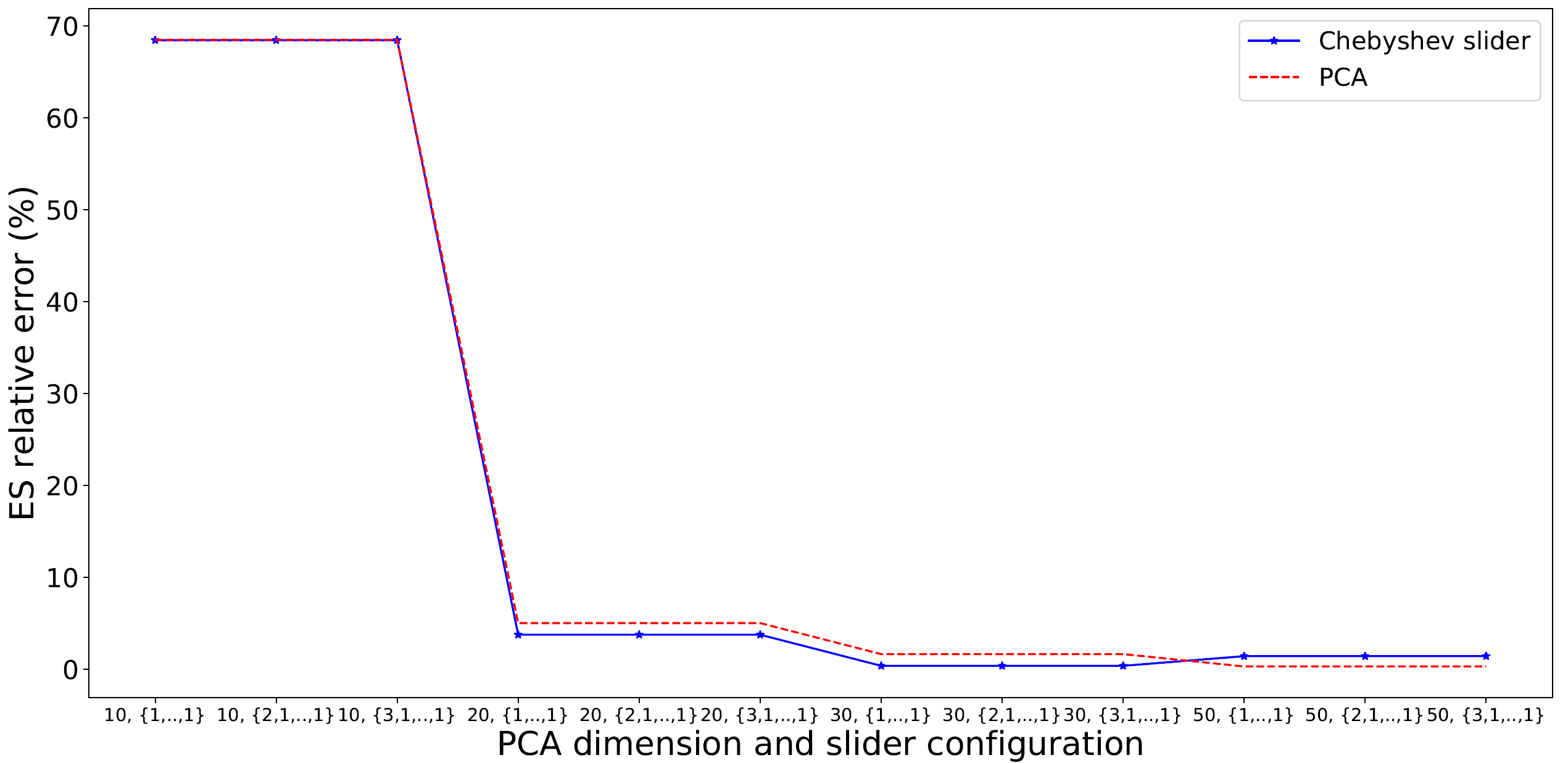}
\caption{This figure shows how the Orthogonal Chebyshev Slider relative ES error changes as dimensionality and slider configuration change for Swaptions $10$-day liquidity horizon.}
\label{fig: plot sens swaptions 10d}
\end{figure}

\begin{figure}
\centering
\includegraphics[scale=0.60]{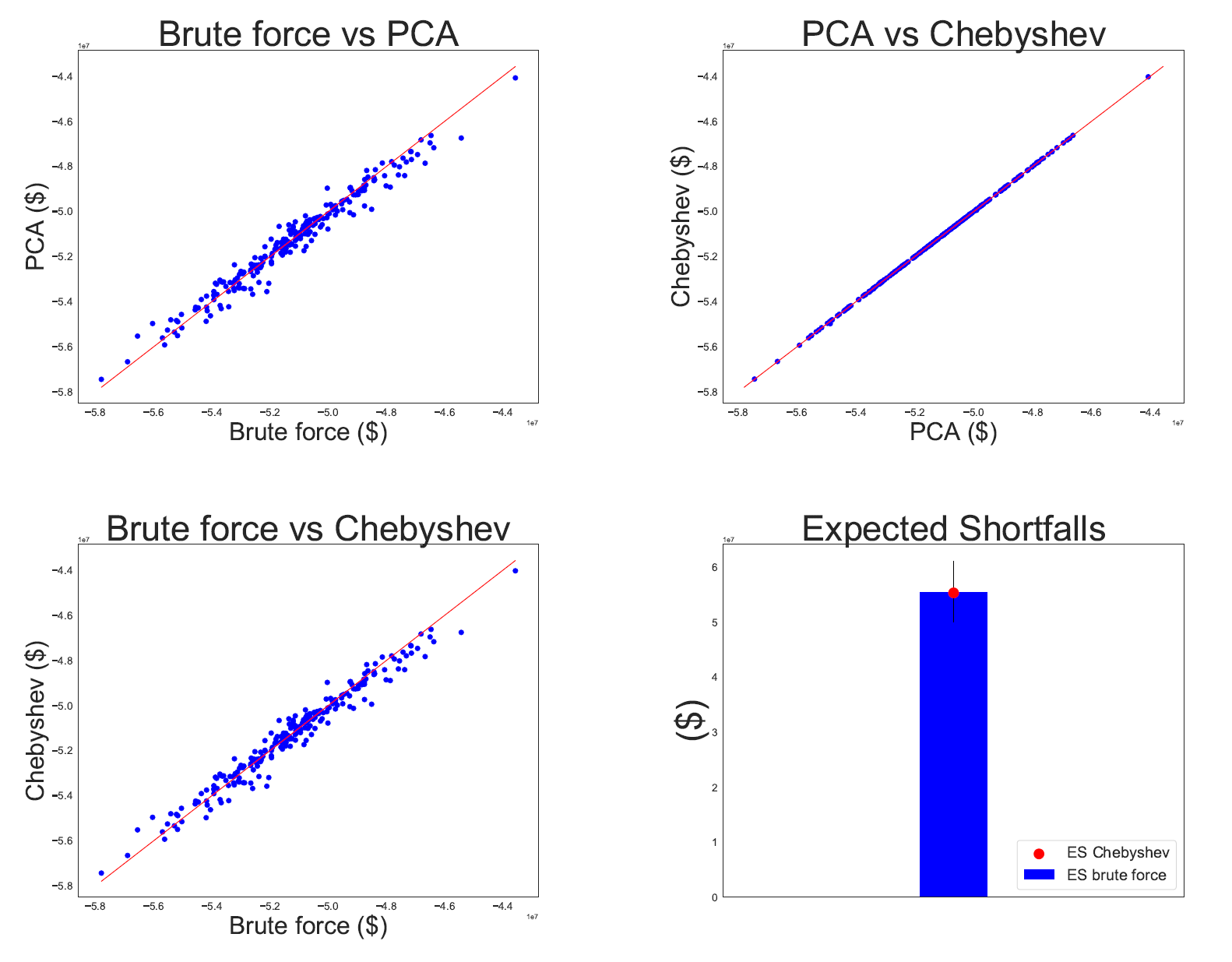}
\caption{Portfolio of Swaptions, slider configuration $\{1,1,\ldots,1\}$, PCA dimension $20$, evaluated on the most recent $250$ scenarios, for $10$-day liquidity horizon.}
\label{fig: plot 250 swaptions 10d}
\end{figure}
\bigskip

Just as with Swaps, the Orthogonal Chebyshev Slider configuration did not make any material difference to the accuracy of the ES.  Figure \ref{fig: plot sens swaptions 10d} shows how there is a big increase in accuracy going from dimension $10$ to $20$ which continues decreasing thereafter. However, as the configuration changes within the same dimension, the accuracy is pretty much unaffected. This is also the case for the $60$-day liquidity horizon as can be seen in Figure \ref{fig: plot sens swaptions 60d}. Note however, that the difference in errors as dimensions vary is not as big in the $60$-day liquidity horizon; even the worse ES relative error, which occurs for dimension $20$, is $2.46\%$, whereas the jump in ES percentage error in the case of the $10$-day liquidity horizon when going from dimension $10$ to $20$, is of more than $60\%$.
\bigskip

The results presented so far for Swaptions in the $10$-day liquidity horizon were obtained for $3,108$ scenarios. If the $20$ dimensional slider with configuration $\{1,\ldots,1\}$ were used on the last $250$ scenarios, for the computation of daily ES within the FRTB IMA framework, one obtains the results shown in Figure \ref{fig: plot 250 swaptions 10d}. The ES relative error is of $0.33\%$ the correlation $0.98$, and Kolmogorov-Smirnov $p$-values $0.93$; all excellent results. However, if the slider had been built on the $250$ scenarios in question, a much lower dimensionality could have been used, reducing the building time considerably. We refer to Remark \ref{rmk: re-usability} for further discussion of this case.
\bigskip

\FloatBarrier

\subsubsection{60-day liquidity horizon}\label{sec: 60-day liq horiz}

\begin{figure}[tp]
\centering
\includegraphics[width=14cm, height=12cm]{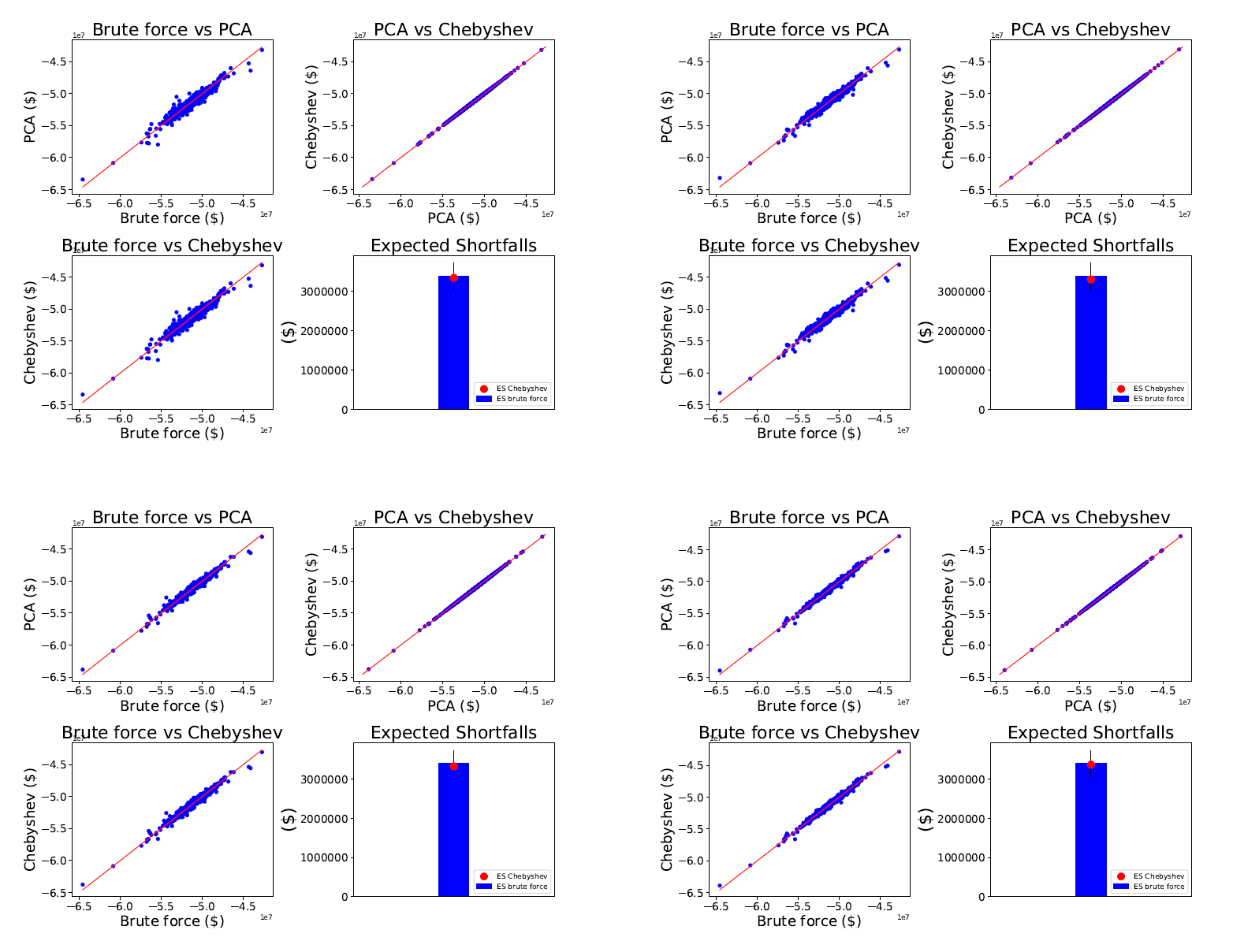}
\caption{Portfolio of Swaptions, slider configuration $\{1,1,\dots,1\}$, on $60$-day liquidity horizon. Top left: PCA dim $10$. Top right: PCA dim $20$. Bottom left: PCA dim $30$. Bottom right: PCA dim $50$.}
\label{fig: plot 3000 swaption 60d}
\end{figure}

\begin{figure}
\centering
\includegraphics[scale=0.45]{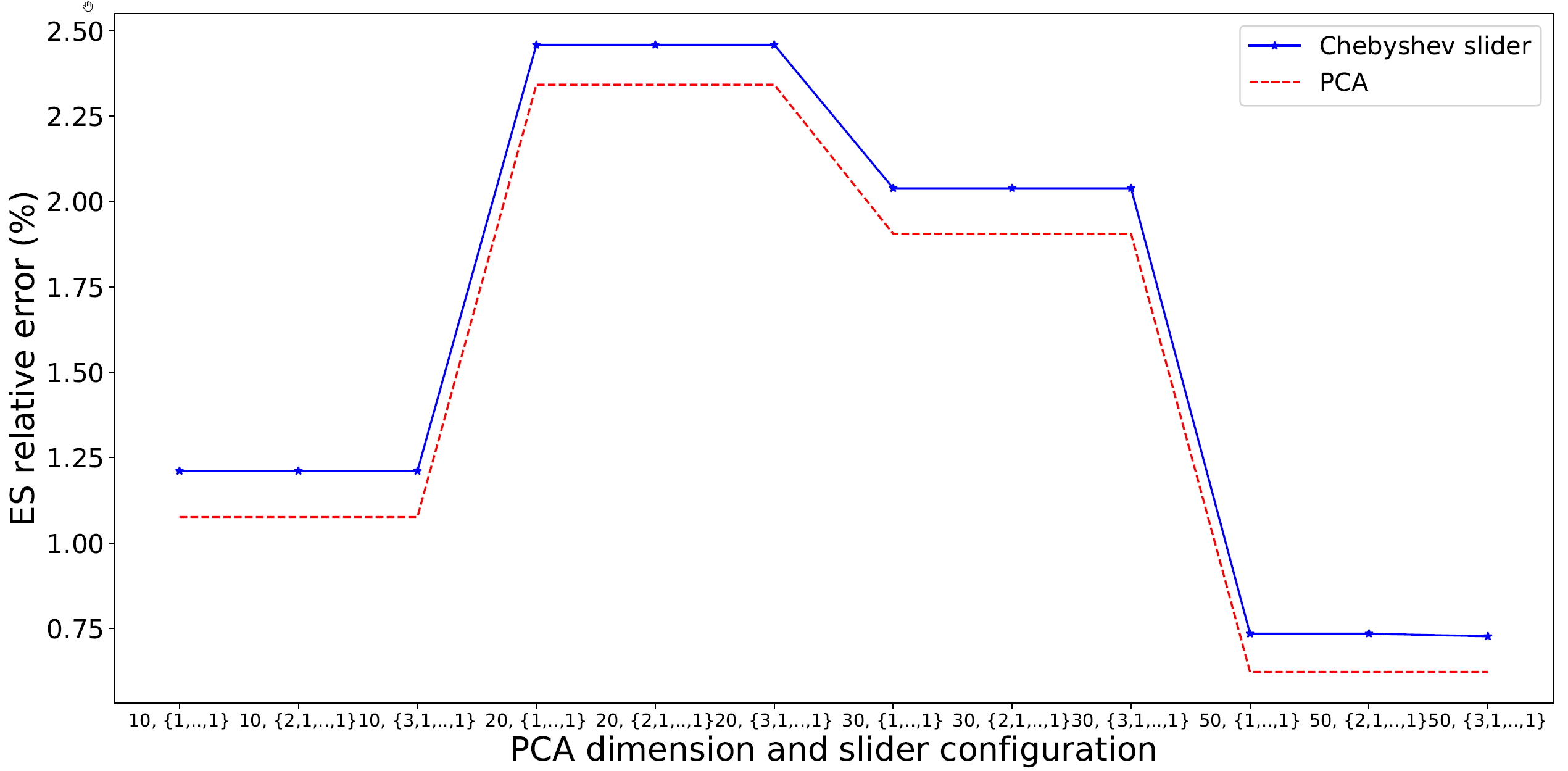}
\caption{This figure shows how the Orthogonal Chebyshev Slider relative ES error changes as dimensionality and slider configuration change for Swaptions $60$-day liquidity horizon.}
\label{fig: plot sens swaptions 60d}
\end{figure}

As explained in Section \ref{sec: diff liq horizons}, the Orthogonal Chebyshev Sliders built for the $10$-day liquidity horizon were used for the $60$-day liquidity horizon. Given that Swaptions are more sensitive to rates and that in the $60$-day liquidity horizon only the volatilities are shocked, the accuracy results are much better. This can be clearly seen in Figure \ref{fig: plot 3000 swaption 60d}, Figure \ref{fig: plot sens swaptions 60d} and Table \ref{tab: swaptions 60d}. Note that as these sliders have already been built for the $10$-day liquidity horizon case, the computational gain is $100\%$.
\bigskip

\begin{figure}
\centering
\includegraphics[scale=0.60]{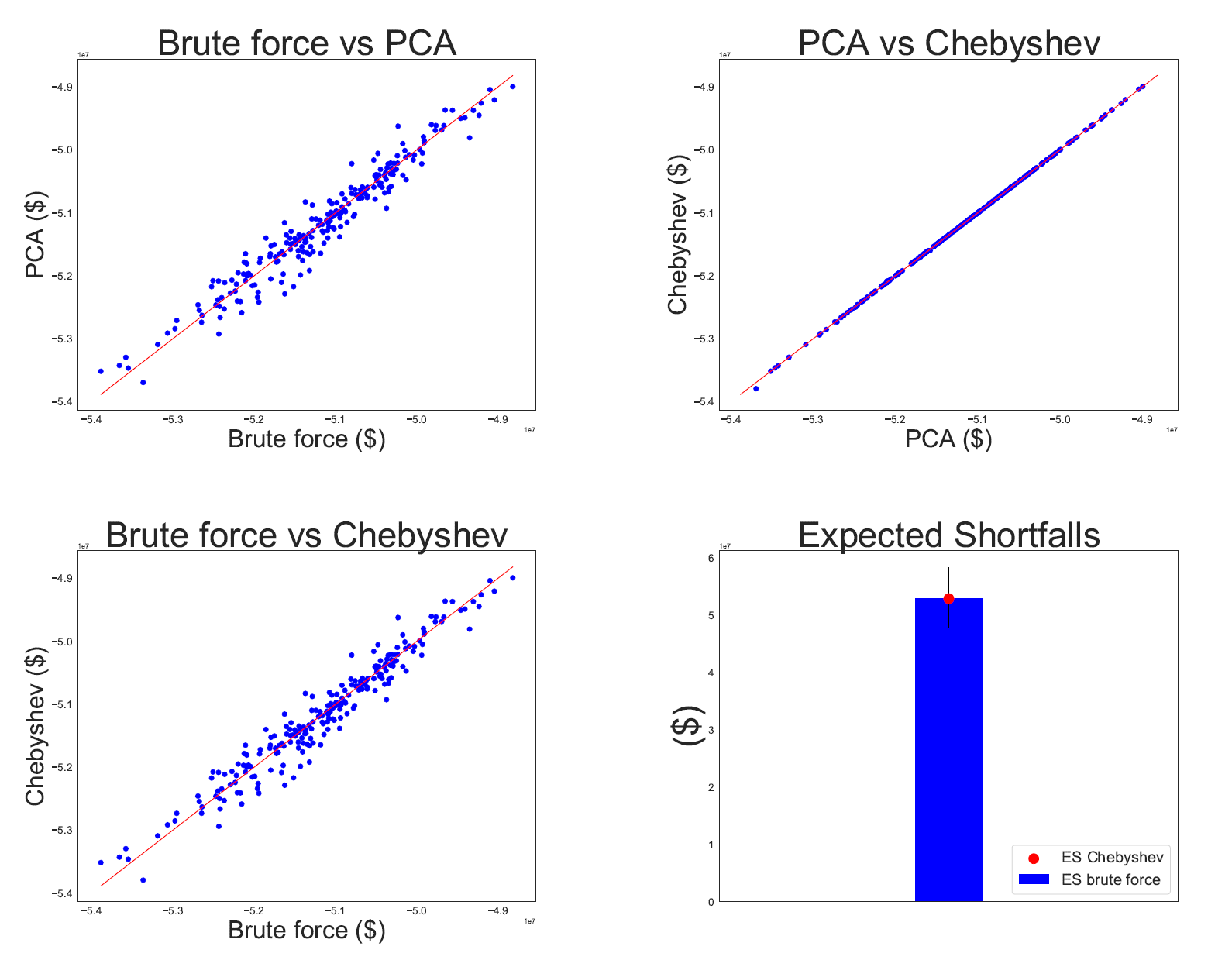}
\caption{Portfolio of Swaptions, slider configuration $\{1,1,\ldots,1\}$, PCA dimension $20$, evaluated on the most recent $250$ scenarios for $60$-day liquidity horizon.}
\label{fig: plot 250 swaptions 60d}
\end{figure}

Just as we have done so far, we evaluate the Orthogonal Chebyshev Sliders on the last $250$ scenarios. The results for the $60$-day liquidity horizon are shown in Figure \ref{fig: plot 250 swaptions 60d}. These correspond to the $20$ dimensional slider with configuration $\{1,\ldots,1\}$. The same one used in the $10$-day liquidity horizon. This time the ES relative error is $0.22\%$, the correlation $1$, and Kolmogorov-Smirnov $p$-value $0.99$; again, excellent results. The computational gain, however, focusing only on the $250$ scenarios, is $80\%$.  As we have discussed, this value would be greatly improved if the Orthogonal Chebyshev Slider was instead built on only the last $250$ scenarios. For a more detailed explanation of this issue we refer to Remark \ref{rmk: re-usability}.
\bigskip

\begin{table}[p]
\centering
\begin{tabular}{|l|l|l|l|l|}
\hline
Slider $\{1, \ldots, 1\}$ & PCA dim $10$ & PCA dim $20$ & PCA dim $30$ & PCA dim $50$\\
\hline
ES relative error & $1.21\%$ & $2.46\%$ & $2.04\%$ & $0.73\%$\\
Computational savings & $100\%$ & $100\%$ & $100\%$ & $100\%$\\
Correlation & $0.97$ &  $0.98$ & $0.99$ & $0.99$\\
KS $p$-value & $0.11$ & $1.00$ & $1.00$ & $0.99$\\
\hline
\end{tabular}
\caption{Portfolio of Swaptions, slider configuration $\{1,1,\ldots,1\}$ on $60$-day liquidity horizon.}
\label{tab: swaptions 60d}
\end{table}

\begin{remark}\label{rmk: re-usability}
The authors are fully aware that the daily computation of ES for FRTB IMA requires a smaller number of evaluations than the ones tested for in this PoC. For Swaps the ES is computed using a distribution of $250$ P$\&$L values while for Swaptions we need two ES values, one for each liquidity horizon, meaning $500$ P$\&$L data points. However, calibrating PCA objects to only $250$ scenarios requires lower dimensions and therefore lower numbers of interpolation points (i.e. calls to the pricing function) which compensates for the lost computational gain when we go from around $3,000$ scenarios down to $250$. As it stands, the Orthogonal Chebyshev Sliders built for $3,000$ scenarios already give savings, on the ES calculation, of $94\%$ for Swaps (dimension $3$ slider) and $80\%$ for Swaptions (dimension $20$ slider), not to mention the possibility of calculating the period of stress which ideally must be done monthly. If we build the Orthogonal Chebyshev Sliders on only $250$ scenarios, we expect to easily increase the computational gains from $80\%$ to $90\%$ in the case of Swaptions.
\end{remark}
\bigskip

In addition to the point made above, the question of re-usability of Orthogonal Chebyshev Sliders must be studied. The scenarios used to obtain the P$\&$L distribution needed for ES calculations change by only a shock on a day to day basis. This means that the PCA built for a Orthogonal Chebyshev Slider on day one will work very well on subsequent days unless there is a big movement in the market. Also, the Orthogonal Chebyshev Slider, essentially a snapshot of the trade on the day its built, will be approximating a trade, on the following day, which will have changed very little: only a day closer to its maturity. These two things can be monitored and if carefully done, one should be able to reuse the Orthogonal Chebyshev Sliders for several days, making the computational savings obtained from using this technique much greater than assumed at first. For example, if the $20$ dimensional Orthogonal Chebyshev Slider with configuration $\{1,\ldots,1\}$ is used for a whole weak ($5$ days), one can expect computational reductions of around $98\%$ in the case of Swaptions.
\bigskip

\begin{remark}
There are two properties of Orthogonal Chebyshev Sliders that makes them re-usable in other calculations with great effect. The first is that their low weight in memory makes serialisation and de-serialisation very efficient. The second is that the evaluation of these objects is extremely fast (see Remark \ref{rmk:barycentric speed}). Therefore, Orthogonal Chebyshev sliders can be used across the organisation in calculations that would be extremely expensive with current risk engines. For example, they can be used in CVA or IMM Monte Carlo simulations. Also in the dynamic versions of these metrics (which usually involve nested Monte Carlo simulations), such as forward capital simulations (KVA, CCAR). Moreover, they can also be used with great effect in accelerating capital optimisation algorithms.
\end{remark}

\FloatBarrier

\section{Conclusion}\label{sec: Key takeaways}

The results presented in this paper were obtained in a PoC that took place within the systems of a tier-one bank. The Expected Shortfall was computed for a portfolio of Swaps and a portfolio of Swaptions, using a set of historic risk factors that spanned around $10$ years (around $3,000$ scenarios). Different liquidity horizons were considered as stipulated within the FRTB IMA framework. The calculations were carried out first, by calling the Front Office pricing functions on every historic shock. These ES values are the benchmark for both accuracy and speed. The aim of the PoC was to test the Chebyshev Sliding Technique (introduced in Section \ref{sec: orth sliding technique}) in terms of accuracy and computational gains by comparing to the benchmark. This is very relevant as the computation of capital calculation with the FRTB IMA framework is computationally very expensive. Therefore, accelerating this calculation is critical for banks both from a performance and business economic standpoint.
\bigskip

The expectations within the PoC were such that an accuracy error of less than $10\%$ and computational gains of more than $50\%$ would be considered a success. As can be seen from Section \ref{sec: Results}, the results far exceeded these thresholds for any of the configurations tested. Out of the three configurations tested, the simplest one ($\{1,\ldots,1\}$) turned out to be the best as it has the highest gains and no material loss of accuracy. For $10$ years of risk factor shocks, the portfolio of Swaps had an error of $0.24\%$ with computational gains of more than $98\%$; in the case of Swaptions, the error was of $3.75\%$ with a computational gain of more than $96\%$ for the $10$-day liquidity horizon, while the $60$-day liquidity horizon (fixed rates, shocked volatilities) had an error of $2.46\%$ with $100\%$ computational gains.\footnote{We remind the reader that the Orthogonal Chebyshev Slider built for the $10$-day liquidity horizon was used in the $60$-day liquidity horizon, hence the $100\%$ computational gain.}
\bigskip

When these Chebyshev Sliders were tested on the last $250$ scenarios, the accuracies were of the same order of magnitude or better, while the computational savings were of about $80\%$. It is important to note, however, as explained in Remark \ref{rmk: re-usability}, that building the Orthogonal Chebyshev Slider on the $250$ scenarios, instead of the $3,000$ they were built on would increase the computational gains up to $90\%$. This was not done as it lied outside the scope of the agreed PoC.
\bigskip

There are two reasons why applying fast and accurate techniques, such as Orthogonal Chebyshev Sliders, to relatively simple products (Swaps and Swaptions), bring big and immediate impacts to the capital calculation of the whole portfolio. First, these products represent a large percentage of the Interest Rate derivative products traded globally (\cite{BIS}) and indeed in a typical portfolio. Although their pricing is generally not considered to be a problem, they account, due to their sheer volume, for a significant part of the computational burden in any risk calculation. Therefore, reducing the computational cost of these products by more than $90\%$, as done in this PoC, brings a big and immediate impact to the computational burden of the whole portfolio. Second, Interest Rate Swaps represent $87\%$ of the gross market value of Interest Rate contracts world-wide (\cite{BIS}). Therefore, accuracy at the level of these products is crucial to obtain accuracy at the level of the whole portfolio.
\bigskip

This PoC was done for portfolios of Swaps and Vanilla Swaptions. However, the authors have also used the techniques presented in this paper on other products like Bermudan Swaptions, within real risk systems, obtaining similar results. The latter should come as no surprise as the set of risk factors for Vanilla Swaptions and Bermudan Swaptions is the same and their pricing function curvatures, qualitatively speaking, very similar. It is important to note that the successful application of Orthogonal Chebyshev Sliders to Bermudan Swaptions has tremendous consequences, as Bermudan Swaptions account, due to their volume in a typical portfolio and high pricing cost, for a good proportion of the computational time of an Interest Rate portfolio.
\bigskip

Another point to examine is the stability of the technique. In this respect, the central questions are if the Orthogonal Chebyshev Sliding Technique will work with stressed market conditions and how often the sliders need to be built. The second is particularly relevant as this is what drives the bulk of the computational burden. The evidence collected so far (see Appendix \ref{sec: Appendix stability}) shows that the stability of the combination of PCA with Chebyshev Sliders is remarkable, hence making it a market turbulence proof technique. Also, evidence suggests that for capital calculations, the building of the sliders does not need to be done daily, at least during periods of normal market conditions. 
\bigskip

\subsection*{Final note}

We would like to thank the reviewer for relevant feedback. Also, please note that some of the techniques presented here fall under the scope of a patent. However, the patent holders are happy to provide licences to anyone interested in implementing them. Please contact the authors for further information.

\pagebreak

\section{Appendix}\label{sec: Appendix}

\subsection{Results for other slider configurations}\label{sec: Appendix diff configs results}

This Section of the Appendix contains the results obtained for the remaining two Orthogonal Chebyshev Slider configurations. That is, the $\{2,1,\ldots,1\}$ and $\{3,1,\ldots,1\}$ configurations, for both Swaps and Swaptions. In the case of Swaptions we have the results corresponding to the $10$-day liquidity horizon and the $60$-day liquidity horizon.

\subsubsection{Swaps}
\FloatBarrier

\begin{table}[H]
\centering
\begin{tabular}{|l|l|l|l|l|}
\hline
Slider $\{2,1, \ldots, 1\}$ & PCA dim $3$ & PCA dim $5$ & PCA dim $10$ & PCA dim $20$\\
\hline
ES relative error & $2.22\%$ & $1.27\%$	& $0.25\%$	& $0.01\%$\\
Computational savings & $99.23\%$ & $98.69\%$ & $97.89\%$ & $96.30\%$\\
Correlation & $0.99$ & $1.00$ & $1.00$ & $1.00$\\
KS $p$-value & $0.85$ & $0.94$ & $1.00$ & $1.00$\\
\hline
\end{tabular}
\caption{Portfolio of Swaps, slider configuration $\{2,1,\ldots,1\}$.}
\label{tab: swaps 2s}
\end{table}

\begin{table}[H]
\centering
\begin{tabular}{|l|l|l|l|l|}
\hline
Slider $\{3,1, \ldots, 1\}$ & PCA dim $3$ & PCA dim $5$ & PCA dim $10$ & PCA dim $20$\\
\hline
ES relative error & $2.23\%$ & $1.30\%$	& $0.25\%$ & $0.00\%$\\
Computational savings & $98.05\%$ & $95.66\%$ & $94.86\%$ & $93.26\%$\\
Correlation & $0.99$ & $1.00$ & $1.00$ & $1.00$\\
KS $p$-value & $0.85$ & $0.94$ & $1.00$ & $1.00$\\
\hline
\end{tabular}
\caption{Portfolio of Swaps, slider configuration $\{3,1,\ldots,1\}$.}
\label{tab: swaps 3s}
\end{table}

\FloatBarrier

\subsubsection{Swaptions}

\FloatBarrier

\begin{figure}[H]
\centering
\includegraphics[width=14cm, height=12cm]{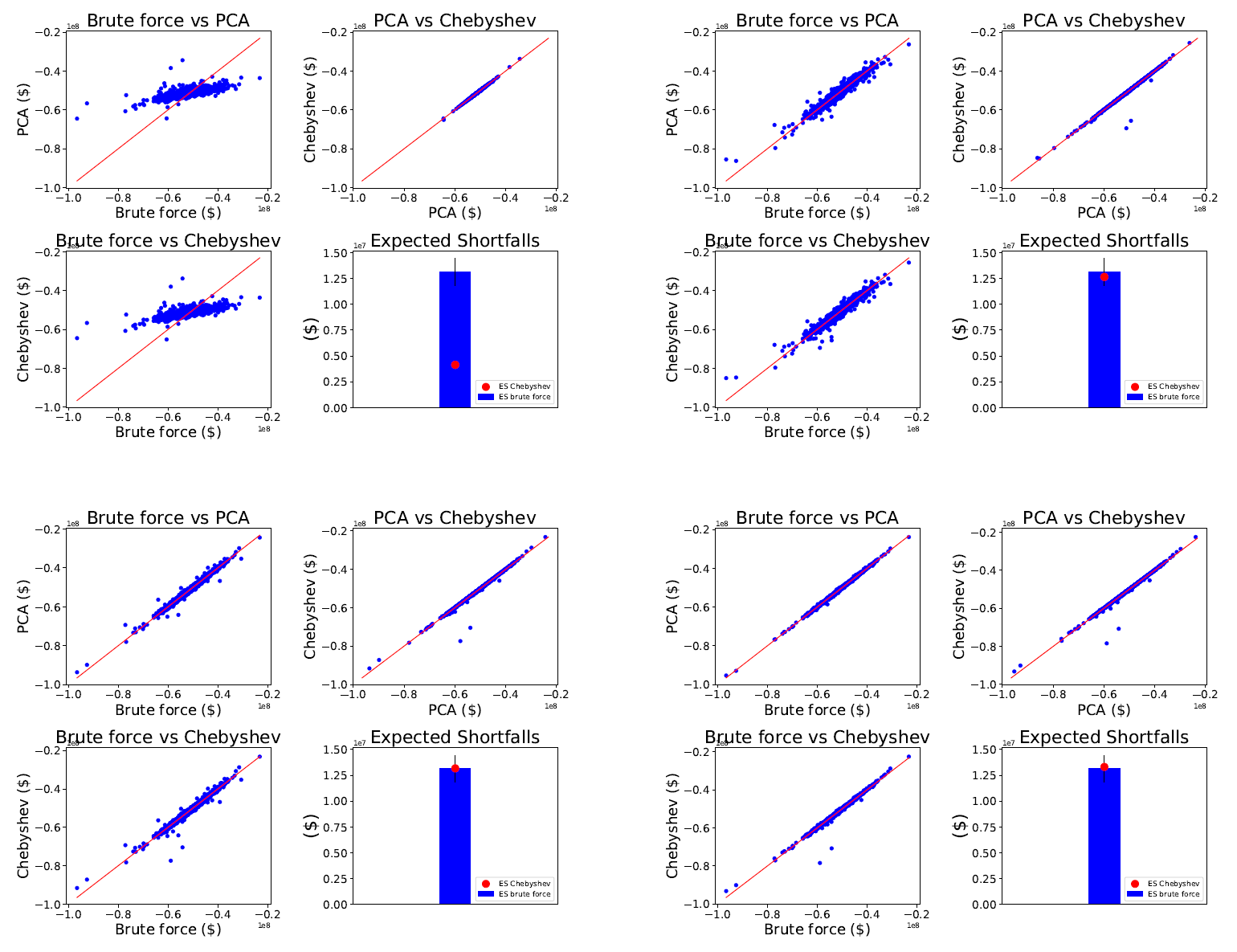}
\caption{Portfolio of Swaptions, slider configuration $\{2,1,\dots,1\}$, on $10$-day liquidity horizon. Top left: PCA dim $10$. Top right: PCA dim $20$. Bottom left: PCA dim $30$. Bottom right: PCA dim $50$.}
\label{fig: plot 3000 swaption 10d 2s}
\end{figure}

\begin{table}[H]
\centering
\begin{tabular}{|l|l|l|l|l|}
\hline
Slider $\{2,1 \ldots, 1\}$ & PCA dim $10$ & PCA dim $20$ & PCA dim $30$ & PCA dim $50$\\
\hline
ES relative error & $68.47\%$ & $3.75\%$ & $0.36\%$	& $1.41\%$\\
Computational savings & $97.88\%$ & $96.27\%$ & $94.66\%$ & $91.44\%$\\
Correlation & $0.75$ & $0.97$ & $0.99$ & $0.99$\\
KS $p$-value & $0.00$ & $0.93$ & $0.83$ & $1.00$\\
\hline
\end{tabular}
\caption{Portfolio of Swaptions, slider configuration $\{2,1,\ldots,1\}$ on $10$-day liquidity horizon.}
\label{tab: swaptions 10d 2s}
\end{table}

\begin{figure}[H]
\centering
\includegraphics[width=14cm, height=12cm]{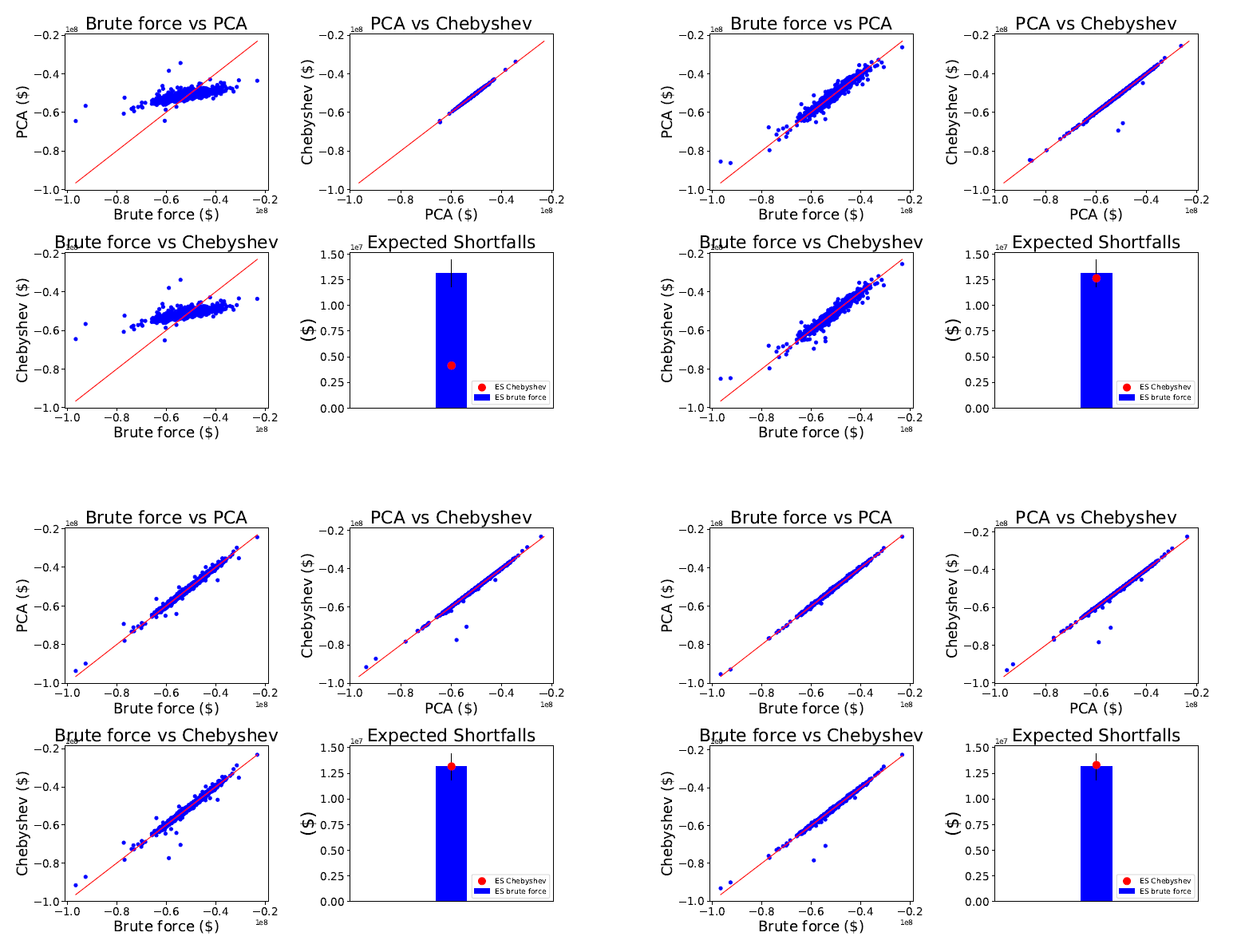}
\caption{Portfolio of Swaptions, slider configuration $\{3,1,\dots,1\}$, on $10$-day liquidity horizon. Top left: PCA dim $10$. Top right: PCA dim $20$. Bottom left: PCA dim $30$. Bottom right: PCA dim $50$.}
\label{fig: plot 3000 swaption 10d 3s}
\end{figure}

\begin{table}[H]
\centering
\begin{tabular}{|l|l|l|l|l|}
\hline
Slider $\{3,1 \ldots, 1\}$ & PCA dim $10$ & PCA dim $20$ & PCA dim $30$ & PCA dim $50$\\
\hline
ES relative error & $68.47\%$ & $3.75\%$ & $0.36\%$	& $1.41\%$\\
Computational savings & $94.82\%$ & $93.21\%$ & $91.60\%$ & $90.48\%$\\
Correlation & $0.75$ & $0.97$ & $0.99$ & $0.99$\\
KS $p$-value & $0.00$ & $0.93$ & $0.83$ & $1.00$\\
\hline
\end{tabular}
\caption{Portfolio of Swaptions, slider configuration $\{3,1,\ldots,1\}$ on $10$-day liquidity horizon.}
\label{tab: swaptions 10d 3s}
\end{table}

\begin{figure}[H]
\centering
\includegraphics[width=14cm, height=12cm]{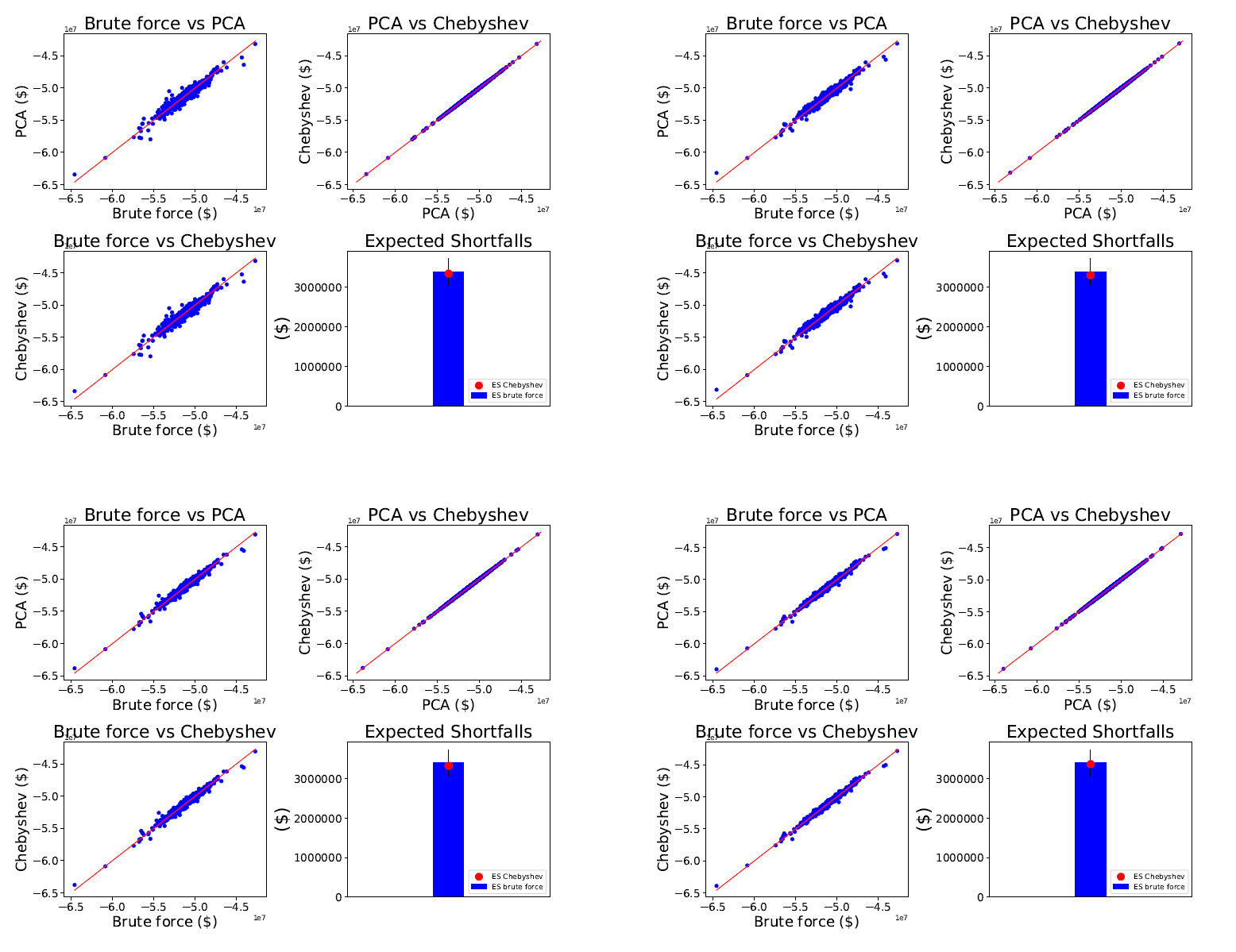}
\caption{Portfolio of Swaptions, slider configuration $\{2,1,\dots,1\}$, on $60$-day liquidity horizon. Top left: PCA dim $10$. Top right: PCA dim $20$. Bottom left: PCA dim $30$. Bottom right: PCA dim $50$.}
\label{fig: plot 3000 swaption 60d 2s}
\end{figure}

\begin{table}[H]
\centering
\begin{tabular}{|l|l|l|l|l|}
\hline
Slider $\{2, \ldots, 1\}$ & PCA dim $10$ & PCA dim $20$ & PCA dim $30$ & PCA dim $50$\\
\hline
ES relative error & $1.21\%$ & $2.46\%$ & $2.04\%$ & $0.73\%$\\
Computational savings & $100\%$ & $100\%$ & $100\%$ & $100\%$\\
Correlation & $1.00$ &  $1.00$ & $1.00$ & $1.00$\\
KS $p$-value & $0.11$ & $1.00$ & $1.00$ & $0.99$\\
\hline
\end{tabular}
\caption{Portfolio of Swaptions, slider configuration $\{2,1,\ldots,1\}$ on $60$-day liquidity horizon.}
\label{tab: swaptions 60d 2s}
\end{table}

\begin{figure}[H]
\centering
\includegraphics[width=14cm, height=12cm]{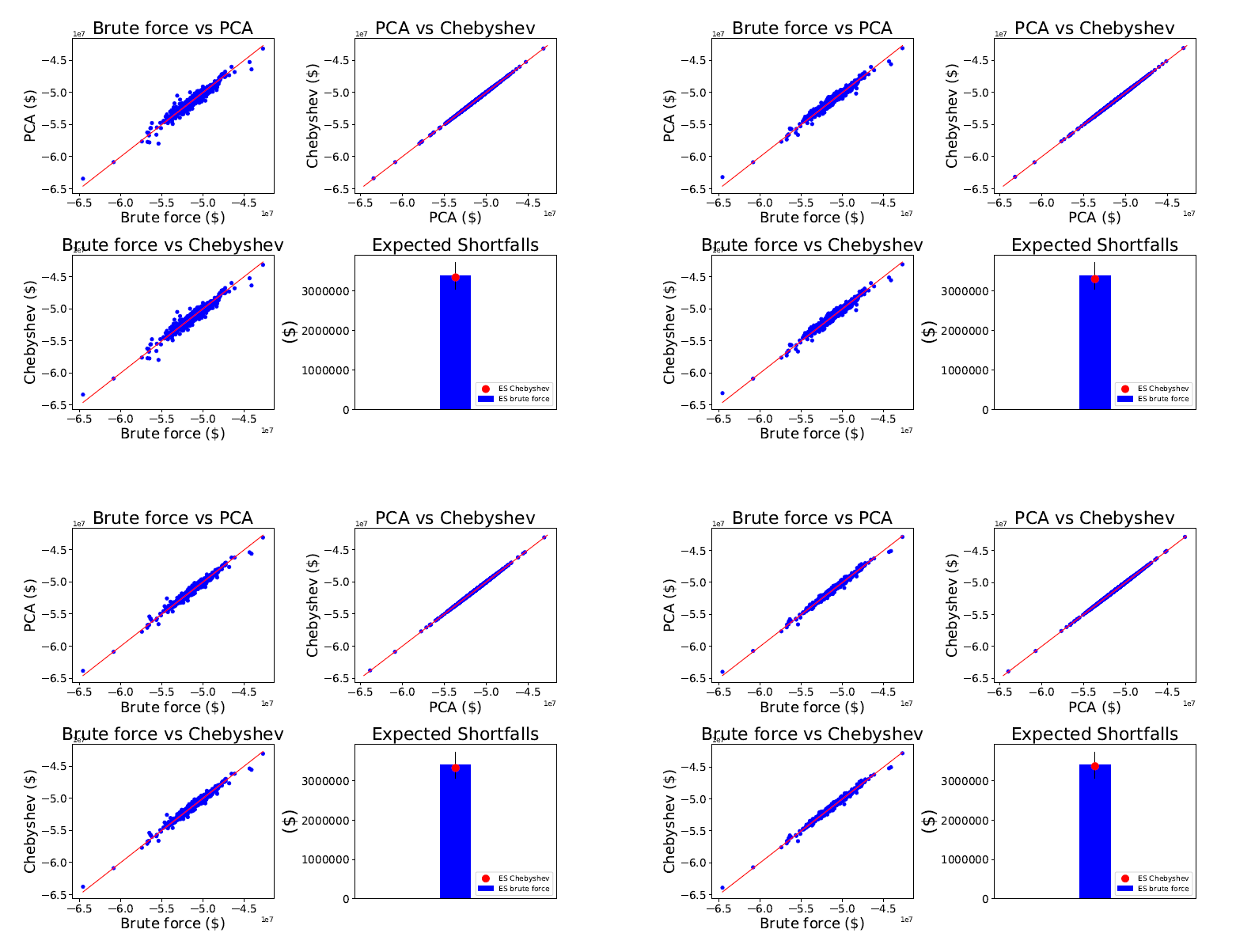}
\caption{Portfolio of Swaptions, slider configuration $\{3,1,\dots,1\}$, on $60$-day liquidity horizon. Top left: PCA dim $10$. Top right: PCA dim $20$. Bottom left: PCA dim $30$. Bottom right: PCA dim $50$.}
\label{fig: plot 3000 swaption 60d 3s}
\end{figure}

\begin{table}[H]
\centering
\begin{tabular}{|l|l|l|l|l|}
\hline
Slider $\{3, \ldots, 1\}$ & PCA dim $10$ & PCA dim $20$ & PCA dim $30$ & PCA dim $50$\\
\hline
ES relative error & $1.21\%$ & $2.46\%$ & $2.04\%$ & $0.73\%$\\
Computational savings & $100\%$ & $100\%$ & $100\%$ & $100\%$\\
Correlation & $1.00$ &  $1.00$ & $1.00$ & $1.00$\\
KS $p$-value & $0.11$ & $1.00$ & $1.00$ & $0.99$\\
\hline
\end{tabular}
\caption{Portfolio of Swaptions, slider configuration $\{3,1,\ldots,1\}$ on $60$-day liquidity horizon.}
\label{tab: swaptions 60d 3s}
\end{table}

\FloatBarrier

\bigskip

\bigskip

\subsection{Results on stability}\label{sec: Appendix stability}
\FloatBarrier

The following results show the stability of Orthogonal Chebyshev Sliders in turbulent markets. The test was done in $2016$ on a portfolio consisting of a variety of Swaps, Bermudan Swaptions, Barrier Options and American Options. The exercise consisted of an out-of-sample back-testing P$\&$L attribution test, on ten years of historical data.\footnote{P$\&$L attribution test as per (\cite{FRTB 2016}) }
\bigskip

\begin{figure}[H]
\centering
\includegraphics[scale=0.4]{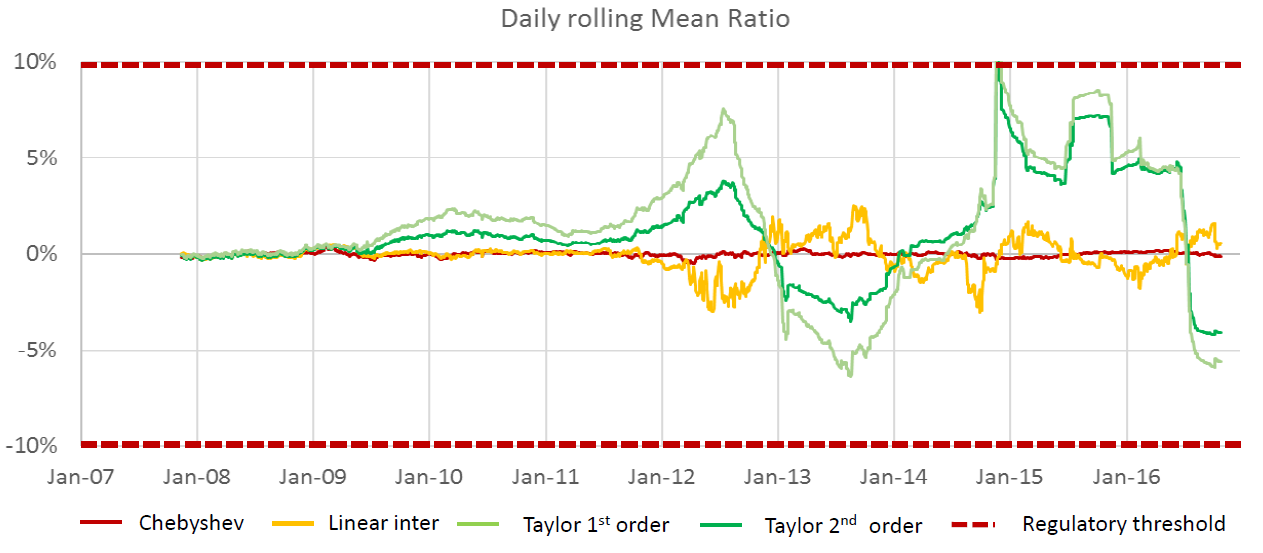}
\caption{Daily rolling mean ratio over a period of ten years for Chebyshev tensor, linear interpolation, and Taylor approximation to first and second order.}
\label{fig: stability mean}
\end{figure}

\begin{figure}[H]
\centering
\includegraphics[scale=0.4]{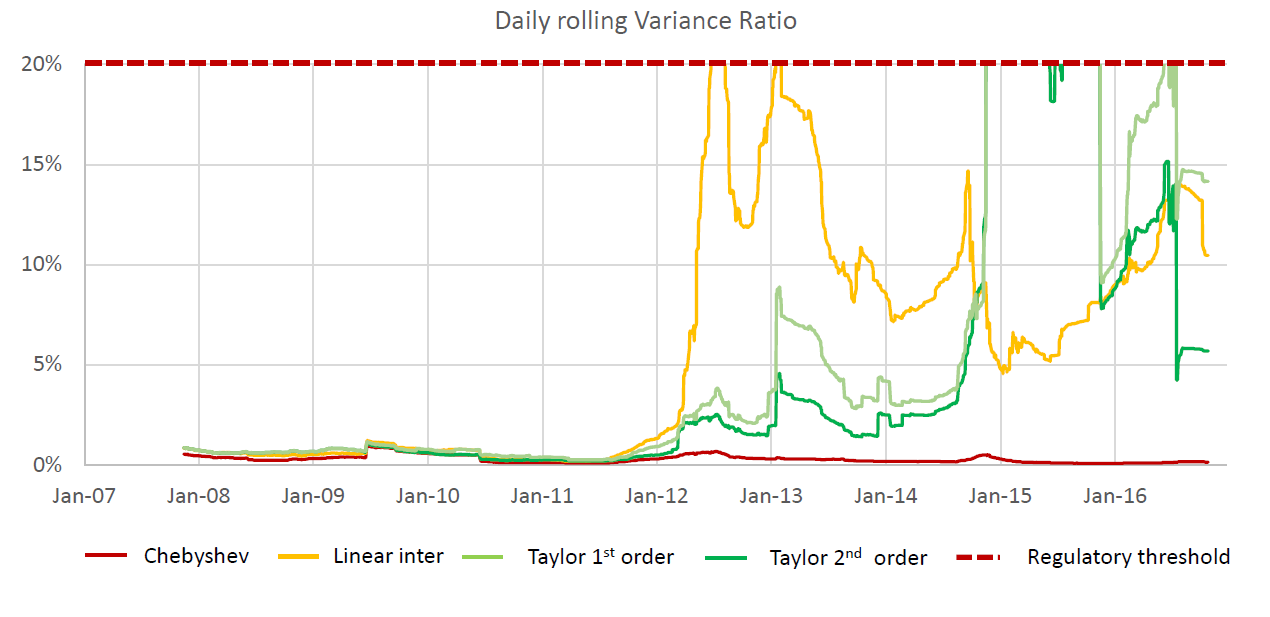}
\caption{Daily rolling variance ratio over a period of ten years for Chebyshev tensor, linear interpolation, and Taylor approximation to first and second order.}
\label{fig: stability variance}
\end{figure}

The hypothetical $P\&L$ was computed using the pricing functions of the trades in the portfolio. The risk theoretical $P\&L$ with one of the following methods: Orthogonal Chebyshev tensors, Tensors on equidistant grids, Taylor expansion (first and second order). Given that four different approximation techniques were used, four different mean ratios and variance ratios time series were obtained. Figure \ref{fig: stability mean} shows the results of all four techniques for the mean ratio test. Figure \ref{fig: stability variance} shows the corresponding time series for the variance ratio tests.

\bigskip

In the first few years, the portfolio was out of the money and its linear behaviour dominated. As a result, all approximation techniques do a good job. In $2012$ (at the peak of the European Sovereign crisis), markets moved substantially and the non-linearities of the portfolio crystallised. As the Figures show, the only technique that was able to price the portfolio with high accuracy, due to the robust mathematical framework it is based on, was the Orthogonal Chebyshev Slider.

\FloatBarrier

\end{document}